\documentclass[a4paper,fleqn,usenatbib]{mnras}

\usepackage{newtxtext,newtxmath}

\usepackage[T1]{fontenc}
\usepackage{ae,aecompl}

\usepackage{graphicx}	
\usepackage{amsmath}	
\usepackage{amssymb}	
\usepackage{pdflscape}
\usepackage{array}

\usepackage{xspace}
\newcommand{\geu}{iPTF16geu\xspace}
\newcommand{\snia}{SN~Ia\xspace}
\newcommand{\sneia}{SNe~Ia\xspace}
\newcommand{\be}{\begin{equation}}
\newcommand{\ee}{\end{equation}}
\newcommand{\om}{\Omega_{\rm M}}

\newcommand{\dl}{D_{\rm l}}
\newcommand{\ds}{D_{\rm s}}
\newcommand{\dls}{D_{\rm ls}}
\newcommand{\zl}{z_{\rm l}}
\newcommand{\zs}{z_{\rm s}}
\newcommand{\te}{\theta_{\rm E}}
\DeclareMathOperator{\tr}{Tr}

\title[Lens modelling iPTF16geu]{Lens modelling of the strongly lensed Type Ia supernova iPTF16geu}
\author[M\"ortsell et al.]{%
E.~M\"ortsell,$^{1}$\thanks{E-mail: edvard@fysik.su.se}
J.~Johansson,$^{2}$
S.~Dhawan,$^{1}$ 
A.~Goobar,$^{1}$
R. Amanullah$^{1}$
D.~A.~Goldstein$^{3,\thanks{Hubble Fellow.}}$
\\
$^{1}$Oskar Klein Centre, Department of Physics, Stockholm University, SE 106 91 Stockholm,
Sweden\\
$^{2}$Department of Physics and Astronomy, Division of Astronomy and Space Physics, Uppsala University, Box 516, SE 751 20 Uppsala, Sweden\\
$^{3}$Division of Physics, Mathematics and Astronomy, California Institute of Technology, Pasadena, CA 91125, USA
}

\date{Accepted XXX. Received YYY; in original form ZZZ}

\pubyear{2019}

\begin{document}
\label{firstpage}
\pagerange{\pageref{firstpage}--\pageref{lastpage}}
\maketitle

\begin{abstract}
In 2016, the first strongly lensed Type Ia supernova, \geu at redshift $z=0.409$ with four resolved images arranged symmetrically around the lens galaxy at $z=0.2163$, was discovered.
Here, refined observations of \geu, including the time delay between images, are used to decrease uncertainties in the lens model, including the the slope of the projected surface density of the lens galaxy, $\Sigma\propto r^{1-\eta}$, and to constrain the universal expansion rate $H_0$.  
Imaging with HST provides an upper limit on the slope $\eta$, in slight tension with the steeper density profiles indicated by imaging with Keck after \geu had faded, potentially due to dust extinction not corrected for in host galaxy imaging.
Since smaller $\eta$ implies larger magnifications, we take advantage of the standard candle nature of Type Ia supernovae constraining the image magnifications, to obtain an independent constraint of the slope. We find that a smooth lens density fails to explain the \geu fluxes, regardless of the slope, and additional sub-structure lensing is needed. 
The total probability for the smooth halo model combined with star microlensing to explain the \geu image fluxes is maximized at $12\,\%$ for $\eta\sim 1.8$, in excellent agreement with Keck high spatial resolution data, and flatter than an isothermal halo. 
It also agrees perfectly with independent constraints on the slope from lens velocity dispersion measurements.
Combining with the observed time delays between the images, we infer a lower bound on the Hubble constant, $H_0 \gtrsim 40\,{\rm km \ s^{-1} Mpc^{-1}}$ at $68.3\,\%$ confidence level.

\end{abstract}




\section{Introduction}
The expansion history of the Universe can be constrained by measuring redshifts and distances of standard candles such as Type~Ia supernovae (\sneia) \citep{2011ARNPS..61..251G}. The redshift gives the growth since the time when the light was emitted. This time is measured, together with the spatial curvature of the Universe, by the distance to the \sneia as inferred from its apparent magnitude.

Weak gravitational lensing from inhomogeneities in the matter distribution will cause a scatter in the distance measurements, possibly degrading the accuracy of the cosmological parameters derived from the expansion history. The first tentative detection of the gravitational magnification of \sneia was made in \cite{Jonsson:2006eu}, see also \cite{Mortsell:2001rb,Jonsson:2005qv,Nordin:2013cfa,Rodney:2015lpa,Rubin:2017ipu}. In principle, the effect can be corrected for by cross-correlating the \sneia observations with data on the foreground galaxies responsible for the lensing effect \citep{Amanullah:2002xh,Jonsson:2005qv,Jonsson:2008xq,Jonsson:2008if}.
The induced scatter can also be used to measure the masses of the foreground galaxies \citep{2010MNRAS.402..526J} and to constrain the fraction of matter inhomogeneities in compact objects \citep{1991ApJ...374...83R,Metcalf:1999qb,Seljak:1999tm,Goliath:1999kpa,Moertsell:2001ah,Mortsell:2001es,Zumalacarregui:2017qqd}, see also \cite{Dhawan:2017kft}.  

In this paper, we study the first resolved strongly lensed \sneia.
It is well known that the time delay between images in strong gravitational lensing systems can be used to constrain the Hubble constant, $H_0$, \citep{1964MNRAS.128..307R}, see also \cite{Goobar:2002pz,Mortsell:2004py,Mortsell:2005mf}. \sneia are especially useful in this respect since \citep{Kolatt:1997zh,Oguri:2002ku}
\begin{enumerate}
\item the time delay can potentially be measured with high accuracy,
\item their standard candle nature can partly break the mass and source sheet degeneracy,
\item their transient nature allows for accurate reference imaging.
\end{enumerate} 
Given the persistent tension [currently at the $4.4\,\sigma$-level \citep{Riess:2019cxk}] between $H_0$ inferred from local distance indicators and the Cosmic Microwave Background (CMB) \citep{Aghanim:2016yuo}, independent measurements of the current expansion rate will shed light on possible explanations \citep[e.g.,][]{Mortsell:2018mfj}.

The first observed \snia with expected multiple images is PS1-10afx at redshift $z=1.388$ and a flux magnification $\mu\sim 30$ \citep{Chornock:2013wj,Quimby:2013lfa}. However, the strong lensing nature of the system was not verified by high spatial resolution imaging. 

Searches for high-redshift core-collapse supernovae, intrinsically fainter than \sneia, has profited from observations through lensing clusters of galaxies  \citep{2009A&A...507...71G,2011ApJ...742L...7A, 2016A&A...594A..54P}.
Four images of the core-collapse supernova Refsdal at $z=1.49$ were also observed in the line of sight of a massive cluster \citep{Kelly:2015vjq}. 
Lens models of the cluster predicted that a fifth image should reappear within 500 days, in agreement with the subsequent detection of the image after $310-380$ days \citep{Kelly:2015xvu}.
Unlike \sneia, lensed core-collapse supernovae can not be used to measure the lensing magnification directly.

The first strongly lensed \snia, \geu at redshift $z=0.409$, was identified by its high magnification $\mu>50$ \citep{Goobar:2016uuf}. Subsequent high spatial resolution imaging confirmed the multiple images of the \sneia.
In \cite{Goobar:2016uuf}, the positions of the SN images with respect to the lensing galaxy were used to construct a lensing model, an isothermal ellipsoid galaxy \citep{1993LIACo..31..571K,1994A&A...284..285K} with ellipticity $\epsilon_e=0.15\pm 0.07$ and mass $M=(1.70 \pm 0.06)\cdot 10^{10}\,M_\odot$ within a radius of $\sim 1\,{\rm kpc}$. The total magnification of the SN images was not well constrained by the model, but
the adopted smooth lens halo predicted brightness differences between the SN images in disagreement with observations, providing evidence for substructures in the lensing galaxy, possibly in the forms of stars. The time delays between images were predicted to be shorter than 35 hours at $99.9\,\%$ confidence level.
The system was also studied in \cite{More:2016sys} where the anomalous image flux ratios were confirmed and time delays between images were predicted to be less than a day. Also, the slope of the lens mass distribution was constrained to be $\eta=2.1\pm 0.1$, consistent with an isothermal profile for which $\eta = 2$. In \citet{Yahalomi:2017ihe}, anomalies between macro-model predictions and the observed \geu flux ratios were investigated, finding that
the discrepancies are too large to be due to microlensing alone. Using the macrolens models from \cite{More:2016sys}, the probability for microlensing to explain the observed flux ratio was found to be $0.3\,\%$. Allowing for alternative macromodels, with varying contributions from external shear and ellipticity, increased the probability to $3\,\%$.

Substantial effort in obtaining follow-up data, including reference imaging after \geu had faded, has resulted in smaller uncertainties in SN image positions and lens galaxy properties, as well as the first observational constraints on the time delay between SN images, as presented in an accompanying paper \citep{Dhawan:2019vof}, summarized in section~\ref{sec:Dhawanetal}.

In this paper, we take advantage of the better observational constraints on the system to refine the lens model and investigate the dependence on the expansion rate $H_0$. We also quantify to what degree lens galaxy stellar microlensing can explain the observed image flux anomalies taking the new data into account.

We use geometrized units for which $c=G=1$ and express the dimensionless Hubble constant, $h$, in units of 100 km/s/Mpc,
\be
h\equiv\frac{H_0}{\rm 100\, km/s/Mpc}.
\ee

\section{Gravitational lensing}
\label{sec:lensing}
When light from a source at angular position $\vec\beta$ and redshift $\zs$, passes a mass at $\zl$ along its path, it is deflected and observed at an angle $\vec\theta$. The time delay, $\Delta t$, compared to an unlensed image is given by
\be\label{eq:timedelay}
\Delta t=\frac{\ds\dl}{\dls}(1+\zl)\tau,\hspace{0.2cm}\tau\equiv\frac{1}{2}(\vec\theta-\vec\beta)^2 - \Psi (\vec\theta),
\ee
where the combination 
\be\label{eq:timedelaydistance}
D_{\Delta t}\equiv \frac{\ds\dl}{\dls}(1+\zl),
\ee
is denoted the time delay distance.
Here, $\dl,\ds$ and $\dls$ are angular distances to the lens, source and between the lens and source, respectively, and
$\Psi (\vec\theta)$ is the scaled, projected Newtonian potential, $\Phi$, of the lens
\be
\Psi (\vec\theta)=2\frac{\dls}{\ds\dl}\int \Phi dl,
\ee 
integrated along the path of the light ray. It is related to the surface mass density of the lens, or convergence through the Poisson equation, $\kappa=\nabla^2\Psi/2$, giving 
\be
\kappa (\vec\theta)=4\pi\frac{\dl\dls}{\ds}\int\rho dl=\frac{\Sigma (\vec\theta)}{\Sigma_{\rm crit}},
\ee
where
\be
\Sigma (\vec\theta)\equiv\int\rho dl\hspace{0.2cm}{\rm and }\hspace{0.2cm}\Sigma_{\rm crit}\equiv \frac{1}{4\pi}\frac{\ds}{\dl\dls}.
\ee
Using Fermat's principle that light rays traverse paths of stationary time with respect to variations of the path, 
we obtain the lens equation 
\be
\vec\beta=\vec\theta-\vec\alpha(\vec\theta),
\ee
where $\vec\alpha(\vec\theta)= \nabla\Psi (\vec\theta)$. Given the Jacobian matrix of the lens mapping,
\be 
\mathcal{A} (\vec\theta)=\frac{\delta\vec\beta}{\delta\vec\theta},
\ee
the magnification, $\mu(\vec\theta)$, is given by 
\be
\mu (\vec\theta)=\frac{1}{\det\mathcal{A} (\vec\theta)}=\frac{1}{(1-\kappa)^2-\gamma^2},
\ee
where the shear $\gamma=\sqrt{\gamma_1^2+\gamma_2^2}$ and (here, subscripts denote partial derivatives with respect to angle components $\theta_i$)
\be
\gamma_1=\frac{1}{2}(\Psi_{11}-\Psi_{22}),\hspace{0.2cm}\gamma_2=\Psi_{12}=\Psi_{21}.
\ee

\subsection{Mass sheet degeneracy}
If we rescale $\Psi$ using the scalars $\xi$ and $u$, and the vector $\vec s$ as
\be
\Psi'=\xi\Psi +\frac{1-\xi}{2}\theta^2+\vec s\cdot\vec\theta+u,
\ee 
the deflection angle $\vec\alpha=\nabla\Psi$ rescales as
\be
\vec\alpha'=\xi\alpha+(1-\xi)\vec\theta+\vec s,
\ee
and the convergence according to
\be
1-\kappa'=\xi (1-\kappa).
\ee
If we also rescale the (unobserved) source position $\vec\beta'=\xi\vec\beta-\vec s$,
the image positions will be unchanged. This is the mass sheet degeneracy; given only the observed image positions, we are free to rescale the projected mass\footnote{Possibly restricted by physical considerations, such as keeping the projected mass positive.}. Magnification and time delay predictions, however, will change according to 
\be
\mu'=\frac{\mu}{\xi^2}\hspace{0.2cm}{\rm and }\hspace{0.2cm}\Delta t'=\xi\Delta t.
\ee
From equation~\ref{eq:timedelay}, the inferred $h$ is proportional to the predicted $\Delta t'$ from the lens model, $h'=\xi h$. Error propagation gives
\begin{align}
\frac{\delta\mu'}{\mu'}&=-2\frac{\delta\xi}{\xi},\\ \nonumber
\frac{\delta h'}{h'}&=\frac{\delta\xi}{\xi}=-\frac{\delta\mu}{2\mu}\approx -0.46 \delta(\Delta m),
\end{align}
where $\Delta m$ is the magnification expressed in magnitudes.
For a \snia, $|\Delta m|\approx 0.1$ and the fractional uncertainty on $h$ from the mass sheet degeneracy is of order 5\,\%, ignoring additional substructure magnifications, see section~\ref{sec:microlensingprob}.

\section{Lens model}
\label{sec:model}
For a cored isothermal ellipsoid, the convergence is given by \citep{1993LIACo..31..571K,1994A&A...284..285K}
\be
\label{eq:CIE}
\kappa =\frac{b}{2\sqrt{s^2+(1-\epsilon)\theta_1^2+(1+\epsilon)\theta_2^2}}, 
\ee
where $b$ is a mass normalization corresponding to the Einstein radius $\te$ for $s=0$, $s$ is the core radius and $\epsilon$ is related to the minor and major axis ratio $q$ as 
\be
q^2=\frac{1-\epsilon}{1+\epsilon}.
\ee
In terms of the ellipticity, $\epsilon_e$, and eccentricity, $e$,
\begin{align}
\epsilon_e &= 1-q = 1-\sqrt{\frac{1-\epsilon}{1+\epsilon}},\\ \nonumber
e &= \sqrt{1-q^2}=\sqrt{\frac{2\epsilon}{1+\epsilon}}.
\end{align}
We can generalize equation~\ref{eq:CIE} as
\be\label{eq:kappa}
\kappa=\frac{b^{\eta-1}}{2\left[s^2+(1-\epsilon)\theta_1^2+(1+\epsilon)\theta_2^2\right]^{(\eta-1)/2}},
\ee
for which $\te=b\sqrt{(1+q^2)/2q}$. In the core-less circularly symmetric case ($\epsilon=s=0$), equation~\ref{eq:kappa} is a good approximation of the projection of a three-dimensional density profile $\rho\propto r^{-n}$, where $r^2=\theta_1^2+\theta_2^2$, for $\eta =n$.
For a singular isothermal sphere (SIS), $\epsilon=s=0$ and $\eta=n=2$. We refer to its elliptic generalization $\epsilon\neq 0$ as a singular isothermal ellipsoid (SIE). For an SIE lens, the image magnification and convergence are related by $\mu_{\rm SIE}=(1-2\kappa_{\rm SIE})^{-1}$.

In terms of the dimensionless $\tau$, the time delay between two lensed images, I and II, of a single source is 
\begin{align}
\Delta\tau &= \tau_{\rm II}-\tau_{\rm I}=\frac{(\vec\theta_{\rm II}-\vec\beta)^2}{2}-\Psi_{\rm II}-\frac{(\vec\theta_{\rm I}-\vec\beta)^2}{2}+\Psi_{\rm I}=\\ \nonumber
&=\frac{r_{\rm I}^2-r_{\rm II}^2}{2}\left[1+\frac{2(\vec\theta_{\rm II}\cdot\vec\alpha_{\rm II}-\vec\theta_{\rm I}\cdot\vec\alpha_{\rm I}+\Psi_{\rm I}-\Psi_{\rm II})}{r_{\rm I}^2-r_{\rm II}^2}\right].
\end{align}
For a SIS halo, $\Psi=\vec\theta\cdot\vec\alpha$ and
\be
\Delta\tau_{\rm SIS}=\frac{r_{\rm I}^2-r_{\rm II}^2}{2}.
\ee 
Converting to polar coordinates, for two images on opposite sides of the lens, we can gain some analytical insight by Taylor expanding in \citep{Mortsell:2004py}
\be
p \equiv\frac{\Delta r}{<r>}=\frac{2(r_{\rm I}-r_{\rm II})}{r_{\rm I}+r_{\rm II}}\text{ and }\delta \equiv\phi_{\rm I}+\pi-\phi_{\rm II}
\ee
obtaining
\be
\label{eq:deltatau}
\Delta\tau=\Delta\tau_{\rm SIS}(\eta -1)\left[1-\frac{(\eta -2)^2}{12}p^2 -\frac{\eta -2}{4}\delta^2\right].
\ee
For symmetric systems, like \geu, for which $p$ and $\delta$ are small, the time delays are simply given by the difference of the squared image distances from the lens centre, times a factor $(\eta-1)/2$.

Note that the time delay only depends on the slope of the projected surface density at radii between the images \citep{1985ApJ...289L...1F}. The main uncertainty affecting the precision of $\Delta\tau$, and thus $h$, is usually the slope of the surface mass density, $\eta$, in the annulus between the images. 

\section{Summary of observations}
\label{sec:Dhawanetal}
We here summarize the observations, and the derived characteristics of \geu, as described in more detail in \citet{Goobar:2016uuf} and \citet{Dhawan:2019vof}.

For the \geu lens system, the redshifts are $\zl=0.2163$ and $\zs=0.409$,
corresponding to angular distances, assuming flat Planck parameter values of $h=0.678$ and $\om=0.308$ \citep{Ade:2015xua},
\begin{align*}
\dl=745.8\,{\rm Mpc},\;\ds=1157.0\,{\rm Mpc}\; {\rm and}\;\dls=513.2\,{\rm Mpc}.
\end{align*}
Expressing positions in arc seconds [arcsec, ''], the time delay distance in equation~\ref{eq:timedelaydistance}, is given by (again assuming flat Planck parameter) $D_{\Delta t}=57.27\,{\rm days}$.

The discovery of \geu prompted observations with the Hubble Space Telescope (HST) Wide Field Camera~3  using the ultra violet and near-infrared (NIR) channels, as well as with laser guided star adaptive optics (LGS-AO) at the Very Large Telescope (VLT) and Keck \citep{Goobar:2016uuf}.  Subsequent data while \geu was active were collected until it disappeared behind the Sun, as well as after it had faded below the detection limit using HST and LGS-AO NIR observations at the Keck telescope in $J, H$ and $K_S$ bands. For $J$ and $H$, we obtained 9 exposures of 20 seconds each following a dithering pattern. For the $K_S$ band, 18~exposures of 65\,seconds were acquired \citep{Dhawan:2019vof}. Together with the data presented in \cite{Goobar:2016uuf}, this resulted in 3 epochs for $J$ and 2 epochs for the $H$ and $K_S$ bands, respectively. The HST data are summarized in table~1 in \cite{Dhawan:2019vof}.

Multiband photometry for the resolved images of \geu is obtained using a combination of a forward modelling approach and template subtractions. The data described above is used to fit for the global lightcurve parameters, including lightcurve peak and shape, as well as color excess from intrinsic color variations or dimming by dust in the host galaxy. The light curves are also corrected for Milky Way dust. Additionally, the resolved images are used to fit for the time offsets between the four SN images, the SN image positions, as well as extinction in the lensing galaxy for each individual line of sight. Assuming a total-to-selective extinction ratio $R_V=A_V/E(B-V)=2$ in the host and lens galaxies, we find $E(B-V)_{\rm host}=0.29\pm 0.05$ mag (common for all images) and $E(B-V)_{\rm lens}$ for images 1, 2, 3 and 4 of $0.06 \pm 0.08$, $0.17\pm 0.08$, $0.42\pm 0.09$ and $0.94\pm 0.07$ mag, respectively, showing that especially image 4 is heavily extinguished in the lens galaxy.
Combining the intrinsic source luminosity standardized from the light curve shapes and colours, with the light curve peak heights and individual image dust extinctions, the magnification factor for each image can be computed. 

The derived image positions, arrival times and magnifications are constrained to the values given in table~\ref{tab:SNpos} \citep{Dhawan:2019vof}. The total magnification of \geu is $\mu_{\rm tot}=67.8^{+2.6}_{-2.9}$. The corresponding time delay differences,  $\Delta t_{ij}\equiv t_j-t_i$, and magnification ratios, $r_{ij}\equiv \mu_i/\mu_j$, are presented in table~\ref{tab:deltobs}. The fact that data is consistent with zero time delay between images means that we can only hope to be able to put a lower limit on $h$.

\begin{table}
  \centering
  \caption{Observed image positions, arrival times and magnifications of \geu images \citep{Dhawan:2019vof}. The radial coordinate is given in arcsec, the angle $\phi$ in radians North of East and the time delay in days.}
  \label{tab:SNpos}
  \begin{tabular}{|c|c|c|c|c|}
    \hline
    Image & $r$ [''] & $\phi$ & $t_{\rm obs}$ [days]& $\mu_{\rm obs}$ \\
    \hline
1 &  $0.251 \pm 0.001$ & $4.468 \pm 0.002$ &  $0\pm 0.33$ & $35.6\pm 1.1$\\
2 &  $0.324 \pm 0.001$ & $2.679 \pm 0.003$ &  $-0.23\pm 0.99$ & $15.7\pm 1.1$\\
3 & $0.297 \pm 0.002$ & $1.013 \pm 0.006$ &  $-1.43\pm 0.64$ & $7.5\pm 1.1$\\
4 & $0.276 \pm 0.001$ & $5.860 \pm 0.005$ &  $1.36\pm 1.07$ & $9.1\pm 1.1$\\
\hline
  \end{tabular}
\end{table}

\begin{table}
  \centering
  \caption{Observed time delay differences,  $\Delta t_{ij}\equiv t_j-t_i$, and  magnification ratios, $r_{ij}\equiv \mu_i/\mu_j$, between \geu images.}
  \label{tab:deltobs}
  \begin{tabular}{|c|c|c|}
  \hline
    Images ($ij$) \quad & $\Delta t_{ij}$ [days] &  $r_{ij}$\\
    \hline
$12$ & $0.23\pm 1.04$ & $2.27\pm 0.17$\\
$13$ &  $1.43\pm 0.81$ & $4.74\pm 0.74$\\
$14$ &  $-1.36\pm 1.12$ & $3.91\pm 0.50$\\
\hline
  \end{tabular}
\end{table}

\section{Macrolens image modelling}
In this section, we constrain the parameters of the macrolens model given in equation~\ref{eq:CIE} using imaging of the lens and host galaxy light distribution, as well as the \geu image positions. Here, the fluxes of the individual \geu image fluxes are not included in the fits. In subsequent sections, we also take advantage of the standard nature of the supernova source to make use of the \geu image magnifications in the modelling procedure.

\subsection{Lens mass constraints from \geu image positions}
In line with the analysis in \citet{Goobar:2016uuf}, we first use only the SN image positions to constrain the parameters of the lens mass model, including $b$, $\epsilon$, the slope $\eta$, the core size $s$, the orientation of the major axis, the position of the centre of mass of the lensing galaxy, and the corresponding \geu source position, employing the {\tt lensmodel} software package \citep{Keeton:2001sr,Keeton:2001ss}. Since the image fluxes can be subjected to systematics effects such as lensing by substructure and dust extinction not properly corrected for, this represents the most conservative approach to the problem\footnote{In fact, it is not possible to successfully reproduce the observed \geu fluxes with the employed lens model, even when allowing for varying slope and core size, $\eta$ and $s$.}. 

The slope and core size are not well-constrained by the SN positions, since varying $\eta$ and/or $s$ can be compensated for by changing $b$ in a way that keeps the total mass within the images constant. 
When varying the slope of the lens, flux ratios vary little, but the total magnification a lot. We can thus get a wide range of magnifications, even without invoking the mass sheet degeneracy. 

At least in principle, the central surface density can be constrained by the fact that we do not observe a central image of \geu. Close to the centre of the lens, the image magnification is given by 
\be
\mu_{\rm central}\approx \frac{1}{(1-\kappa)^2}.
\ee
If we can observationally constrain $\mu_{\rm central}<\mu_{\rm max}$, then (for $\kappa>1$)
\be\label{eq:sigcent}
\kappa >1+\frac{1}{\sqrt{\mu_{\rm max}}}\rightarrow 
\Sigma_{\rm central}>\Sigma_{\rm crit}\left(1+\frac{1}{\sqrt{\mu_{\rm max}}}\right),
\ee
where $\Sigma_{\rm crit}\approx 10.50\,{\rm kg}/{\rm m}^2$ for \geu. Since we expect the central image to be subject to large dust extinction not accounted for in the analysis, equation~\ref{eq:sigcent} represents the most optimistic limit obtainable for the lens galaxy central surface density. Assuming a Hsiao template \citep{2007ApJ...663.1187H}, with a Bessell B band magnitude of -19.3, reddened by $A_V=0.23$ magnitudes in the Milky Way and $A_V=0.58$ magnitudes in the host galaxy, the $3\,\sigma$ upper limit on a possible fifth central image corresponds to a demagnification of $1.2$ magnitudes, or $\mu_{\rm max}=0.33$, implying $\Sigma_{\rm central}>29\,{\rm kg}/{\rm m}^2$. At least in principle, the lack of a detected central image can be used also to constrain the slope of a parametrized lens mass model. However, in the case of \geu, also in the case of a very flat halo with $\eta=1.2$, the central image would be below our detection limit, even when ignoring dust extinction.

\subsection{Lens mass constraints from imaging data}
In order to constrain also the slope of the deflector, we include full imaging data using {\tt lenstronomy}, an open-source python package using forward modelling to reconstruct strong lensing systems \citep{Birrer:2018xgm}.
We simultaneously reconstruct the lens mass model, the SN images (when applicable) and the surface brightness distributions of the lens and the lensed \geu host galaxy. Compared to systems of strongly lensed quasars, we can take advantage of the fact that the transient nature of SNe allows for accurate reference imaging, i.e., we have a clear view of the lensed host when the SN has faded.

The lens mass is described by equation~\ref{eq:kappa}, using zero core size, $s=0$.
For the lens galaxy and (unlensed) host galaxy light intensity, we assume elliptical Sérsic profiles 
\be
I(R) = I_e\exp{\left[-b_n\left(\frac{R}{R_e}\right)^{1/n}-1\right]},
\ee
where $I_e$ is the intensity at the half-light radius $R_e$, $b_n=1.9992 n -0.3271$ \citep{Birrer:2018xgm} and 
\begin{align}
R\equiv\sqrt{x_1^2+x_2^2/q_S^2},
\end{align}
where $x_1$ and $x_2$ are coordinates aligned with the major and minor axes of the galaxy light distribution and $q_S$ is the axis ratio.

\subsubsection{HST imaging data}
We expect to get the most reliable and conservative results from the lens modeling from HST data, given the relatively well known and stable point spread functions (PSFs), determined empirically in Appendix B of \cite{Dhawan:2019vof}. However, due to their limited resolution, compared to the size of the lens system, HST data may not be very effective in constraining the slope $\eta$ of the lens mass distribution.
Out of the available HST filters, we concentrate on $F814W$ data, being least subject to dust extinction and displaying the most evident separation between the SN, the lens galaxy and the host galaxy fluxes.

Minimizing the residual difference between the observed and the model image for a range of lens mass slopes in the interval $1.2<\eta<2.5$, we find that the $F814W$ data only provide an upper limit of $\eta\lesssim 1.7$ at $95\,\%$ confidence level (CL), in contrast to previous work in which the slope was constrained to $\eta=2.1\pm 0.1$ \citep{More:2016sys}. The best fit model, for which $\eta =1.2$ has a reduced chi-square of $\chi_\nu^2=0.99735$. Low values for $\eta$ are also preferred by $F814W$ reference imaging after \geu had faded, although the overall quality of the fit is somewhat decreased compared to the imaging when \geu was active. We note that a possible reason for this, and also for the slightly discrepant value of $\eta$ compared to constraints from Keck imaging, velocity dispersion measurements and \geu image magnifications discussed in upcoming sections, is that the host images are affected by dust absorption as evident from the \geu images. Also, figure~3 in \citet{2020arXiv200410164J} show evidence of colour gradients in the host galaxy along the Einstein ring, possibly related to lens galaxy dust extinction. 
Given the longer integration time of the HST reference images, the noise level is decreased, and the sensitivity to not correcting the host galaxy images for the differential dust extinction (which is beyond the scope of this paper) is increased.

\subsubsection{Keck imaging data}
Since constraining the slope relies on high spatial resolution data, preferably at long wavelengths in order to minimize sensitivity to dust extinction, we also investigate to what extent Keck host imaging data can provide additional information about the lens mass distribution. The Keck LGS-AO PSF may vary substantially both in time and over the focal plane, potentially making the lens modelling less reliable than for HST data, despite the higher resolution of the Keck LGS-AO data. Since \geu and reference imaging in the $H$-band were obtained with different instruments, and conditions during $K_S$-observations were unstable, we will use data obtained in the $J$-band, having the highest quality out of the three Keck bands.

\begin{figure}
    \centering
	\includegraphics[width=\linewidth]{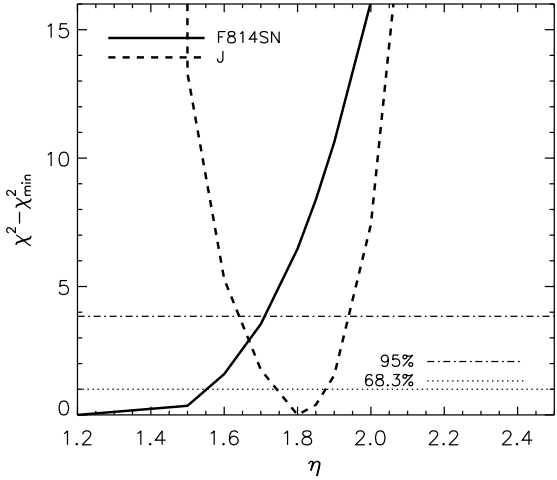}
	\caption{The $\chi^2$ obtained from modelling the \geu system using observations in the HST $F814W$ band when the SN was active (F814SN, solid line) and Keck $J$-band reference imaging (J. dashed line) as a function of the lens mass slope $\eta$. The horizontal dotted and dash-dotted lines indicate $68.3\,\%$ and $95\,\%$ CL, respectively. $F814W$-data indicate $\eta < 1.7$, whereas $J$-data gives $\eta = 1.8\pm 0.15$ at $95\,\%$ CL.
	\label{fig:etaplot}}
\end{figure}

We approximate the PSF as a Moffat profile being constant in time and space. Since there are no isolated stars in the field, we determine its parameters from the SN images, after subtracting the reference image. We then model the lens mass distribution from the reference image of the lensed host galaxy, since we expect this to be less sensitive to the exact shape of the PSF, compared to imaging when \geu is active. The lens model parameters obtained are consistent with that from HST data, except that it favours a slightly higher slope of $\eta = 1.8\pm 0.15$ at $95\,\%$ CL, see figure~\ref{fig:etaplot}. Adding information on the \geu image positions, when fitting the $J$-band image slightly shifts the best fit value of the slope to $\eta\sim 1.7$.

Summarizing, there is a slight tension between the lens mass slope best fitting the HST and Keck data. One reason could be that the assumed mass and light distribution models are too restrictive, another that the tension is caused by incomplete descriptions of the image PSFs. Also, given the large extinction evident in the \geu images, we can not rule out the possibility that the HST host images are affected by dust extinction, not corrected for in the present analysis. In the next section we investigate how the slope can be independently constrained using the observed magnifications of the \geu images.

\begin{figure*}
    \centering
	\includegraphics[width=1\textwidth]{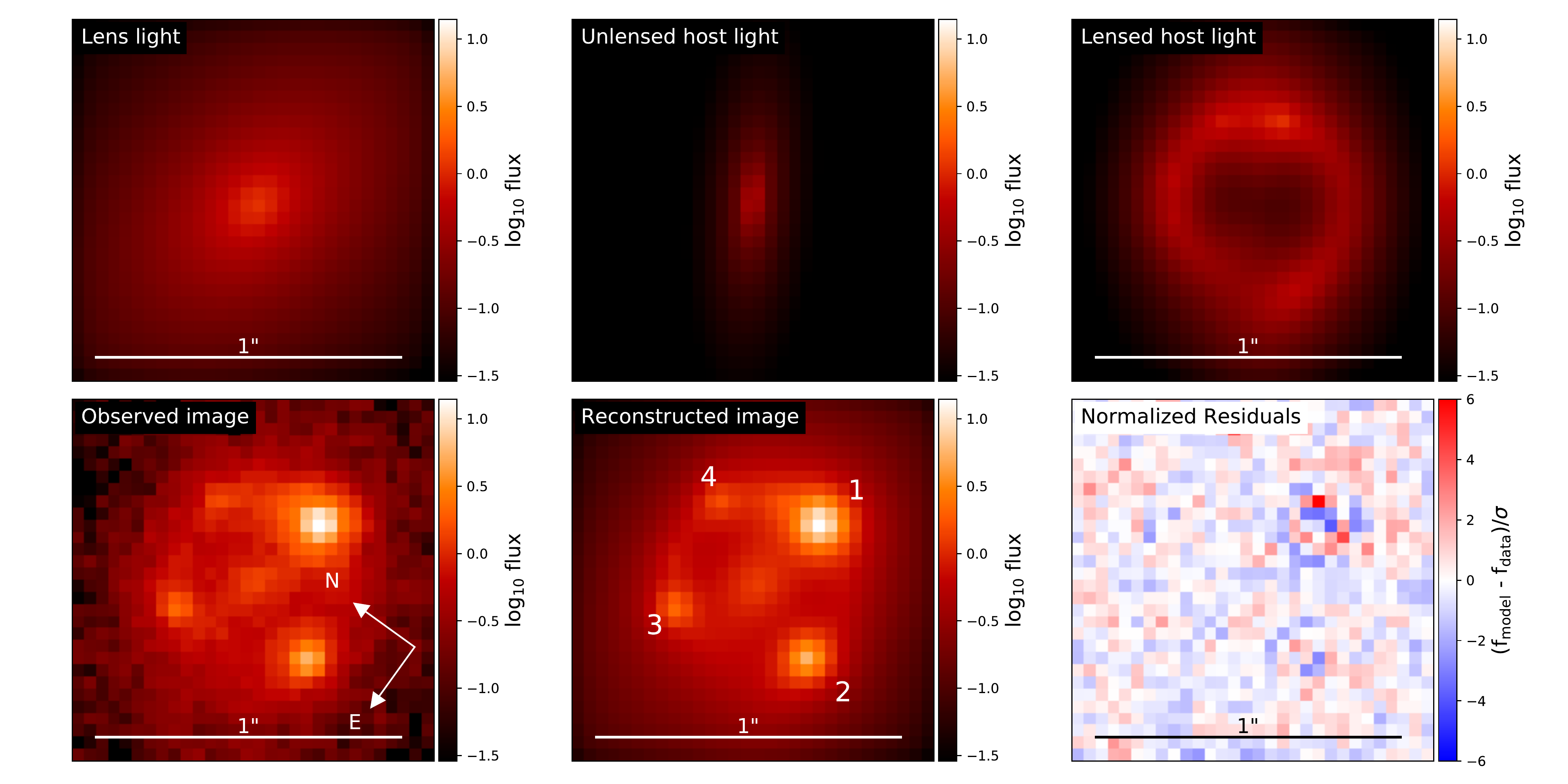}
	\caption{A comparison between the observed (data) image and the reconstructed (model) image for the HST $F814W$-band as of October 25, 2016, assuming a lens mass slope of $\eta=1.8$. {\it Upper panel from left to right:} The reconstructed lens galaxy light intensity, the unlensed host and the lensed host image. {\it Lower panel from left to right:} The observed $F814W$-band image, the total reconstructed image (convolved with the PSF of the instrument, but without background or Poisson noise) and normalized residuals. As indicated by the compass in the lower left panel, the images are rotated with respect to the Keck $J$-band image in figure~\ref{fig:J_init_g18} and \ref{fig:critlines}. In the lower mid panel, the \geu images are labelled according to the numbering given in table~\ref{tab:SNpos}.
	\label{fig:F814_SN_init_g18}}
\end{figure*}

\begin{figure*}
    \centering
	\includegraphics[width=1\textwidth]{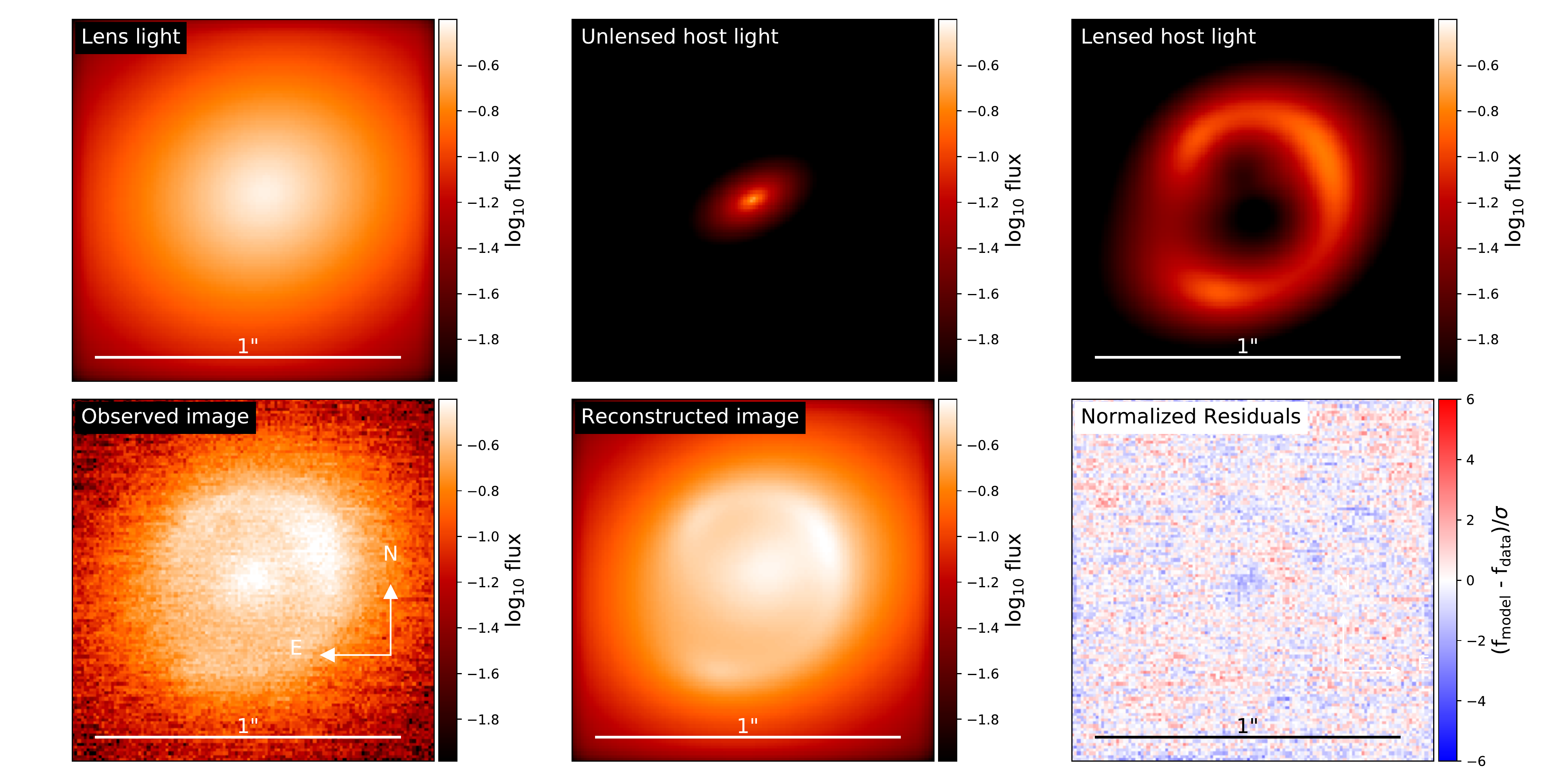}
	\caption{A comparison between the observed (data) image and the reconstructed (model) image for the Keck $J$-band reference image as of June 16, 2017, assuming a lens mass slope of $\eta=1.8$. {\it Upper panel from left to right:} The reconstructed lens galaxy light intensity, the unlensed host and the lensed host image. {\it Lower panel from left to right:} The observed $J$-band image, the total reconstructed image (convolved with the PSF of the instrument, but without background or Poisson noise) and normalized residuals.
	\label{fig:J_init_g18}}
\end{figure*}

Figure~\ref{fig:F814_SN_init_g18} and \ref{fig:J_init_g18} show the observed and reconstructed HST $F814W$-band and Keck $J$-band image for $\eta=1.8$ (for reasons that will soon become evident). We obtain very similar results for lower values of $\eta$, the only exception being that the intrinsic size and luminosity of the host galaxy becomes smaller since the magnification increases as $\eta$ decreases. From left to right, the upper panel shows the reconstructed lens galaxy light distribution, the unlensed host and the lensed host image. In the lower panel, the observed image is shown to the left and in the middle the total reconstructed image, i.e. the sum of the lens and the lensed host light intensity. The reconstructed images have been convolved with the PSF of the instrument, but do not include background or Poisson noise. To the right, normalized residuals after subtracting the reconstructed (model) image and the observed image (data) are shown.

\section{Macrolens velocity dispersion modelling}
An independent velocity dispersion measurement of the lensing galaxy can be used to break the mass sheet degeneracy, and to constrain the slope of the lens mass distribution. \citet{2020arXiv200410164J} present high-resolution spectroscopic observations of iPTF16geu, and derive a velocity dispersion of the lens galaxy, $\sigma_{\rm obs}=129\pm 4\,{\rm km/s}$ using stellar continuum fitting. This value is lower, but consistent with the value derived from emission lines in low-resolution spectra, $\sigma_{\rm obs}=163^{+41}_{-27}\,{\rm km/s}$  \citep{Goobar:2016uuf}.

The theoretical velocity dispersion for a spherically symmetric system can be derived from the equations of stellar hydrodynamics:
\be
\frac{d}{dr}(\nu\sigma_{r}^{2})+\frac{2\beta}{r}\nu\sigma_{t}^{2}=-\nu\frac{d\Phi}{dr}\equiv-\nu\Phi^{\prime},
\ee
where $\sigma_{t}$ and $\sigma_{r}$ are the velocity dispersions in the tangential and radial direction, respectively, $\beta=1-(\sigma_{t}/\sigma_{r})^{2}$ is the velocity anisotropy, $\nu$ is the three dimensional density of velocity dispersion tracers (luminous matter) and $\Phi$ is the total gravitational potential. The prime indicates differentiation with respect to $r$. Assuming constant $\beta$, 
\begin{equation}\label{eq:sigmar2}
\sigma_{r}^{2}(r)=\frac{1}{\nu r^{2\beta}}\intop_{r}^{\infty}\nu x^{2\beta}\Phi^{\prime}dx.
\end{equation}

To estimate the expected velocity dispersion of the \geu lens galaxy, we approximate the light and mass distributions as spherically symmetric and deproject the best-fit Sérsic profile for the lens luminosity distribution and the power-law projected total mass profile to obtain $\nu (r)$ and $\Phi (r)$ respectively\footnote{We have used codes developed by Matthew Auger and Alessandro Sonnenfeld available at \url{https://github.com/astrosonnen/spherical_jeans}.}.

\begin{figure}
    \centering
	\includegraphics[width=\linewidth]{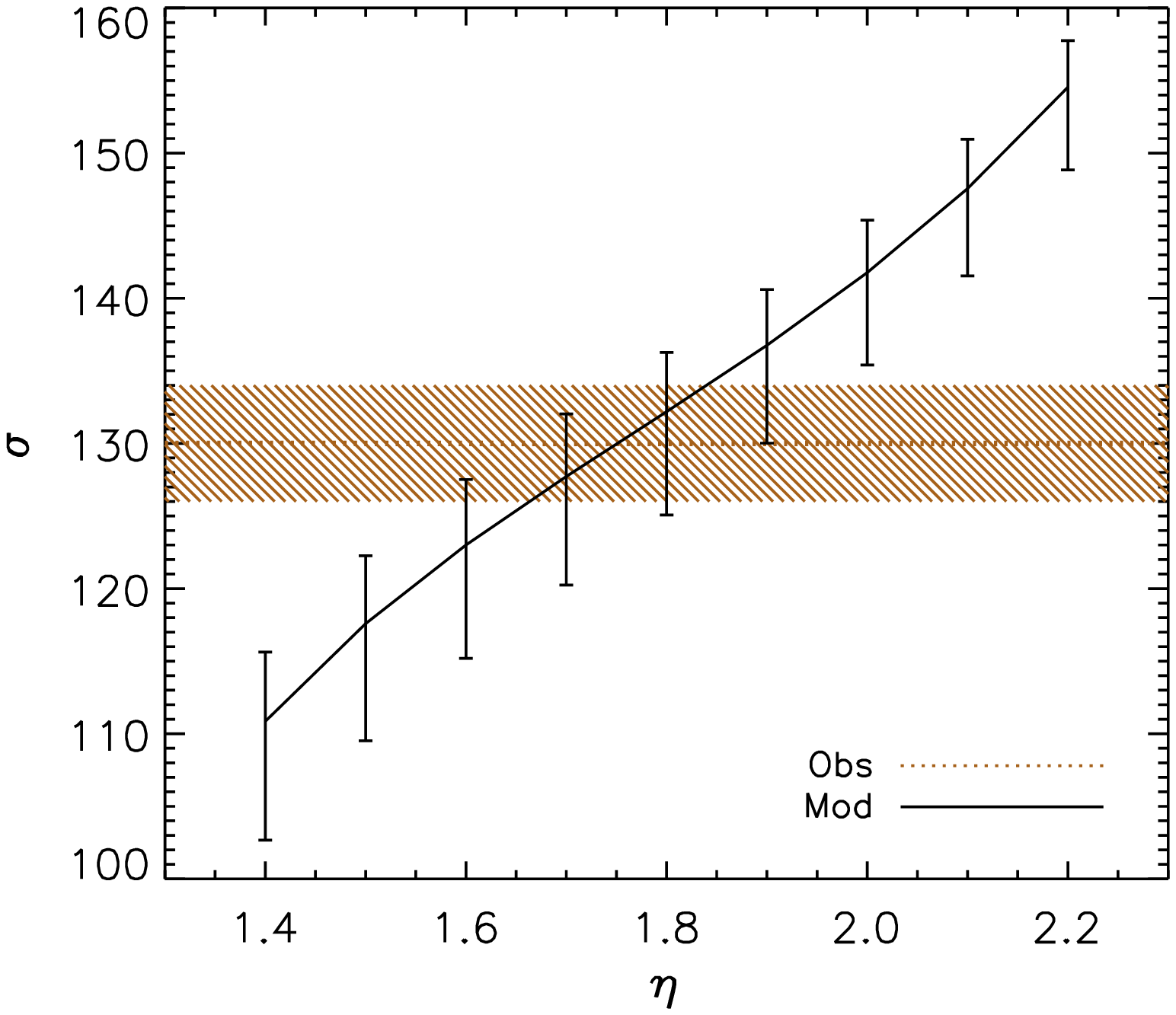}
	\caption{The predicted lens galaxy velocity dispersion as a function of the lens mass slope $\eta$. The value $\eta=2$ represents an isothermal halo. The horizontal bands indicate the observed velocity dispersion of $\sigma_{\rm obs}=129\pm 4\,{\rm km/s}$ \citep{2020arXiv200410164J}, favouring values for the slope in the interval $1.6<\eta<2$.
	\label{fig:vdispplot}}
\end{figure}

The lens luminosity distribution is derived from HST $F475W$, expected to most closely trace the star distribution used when observationally constraining the velocity dispersion, see \citet{2020arXiv200410164J}. In doing this, we keep the lens mass distribution obtained from $F814W$ data fixed.
The model velocity dispersion, $\sigma_{\rm mod}$, is given by a line-of-sight luminosity weighted average over the effective spectroscopic aperture of the observations, here approximated as a circular aperture with radius $0.9$ arcsec. In figure~\ref{fig:vdispplot}, we show $\sigma_{\rm mod}$ as a function of lens mass slope $\eta$. Compared with the observed velocity dispersion of $\sigma_{\rm obs}=129\pm 4\,{\rm km/s}$ indicated by the horizontal band, we see that $1.6<\eta<2$ are prefered, in reassuring agreement with the values obtained from the \geu image fluxes in section~\ref{sec:microlensingprob}.
The main uncertainty is the anisotropy parameter, $\beta$, for which we conservatively assume $\beta = 0.2^{+0.2}_{-0.5}$ \citep{Gerhard:2000ck,2019ApJ...874...41C}, yielding $\sigma_{\rm mod}=132^{+4}_{-7}\,{\rm km/s}$ for $\eta = 1.8$.

\section{Macrolens image magnifications modelling}
\begin{figure}
    \centering
	\includegraphics[width=\linewidth]{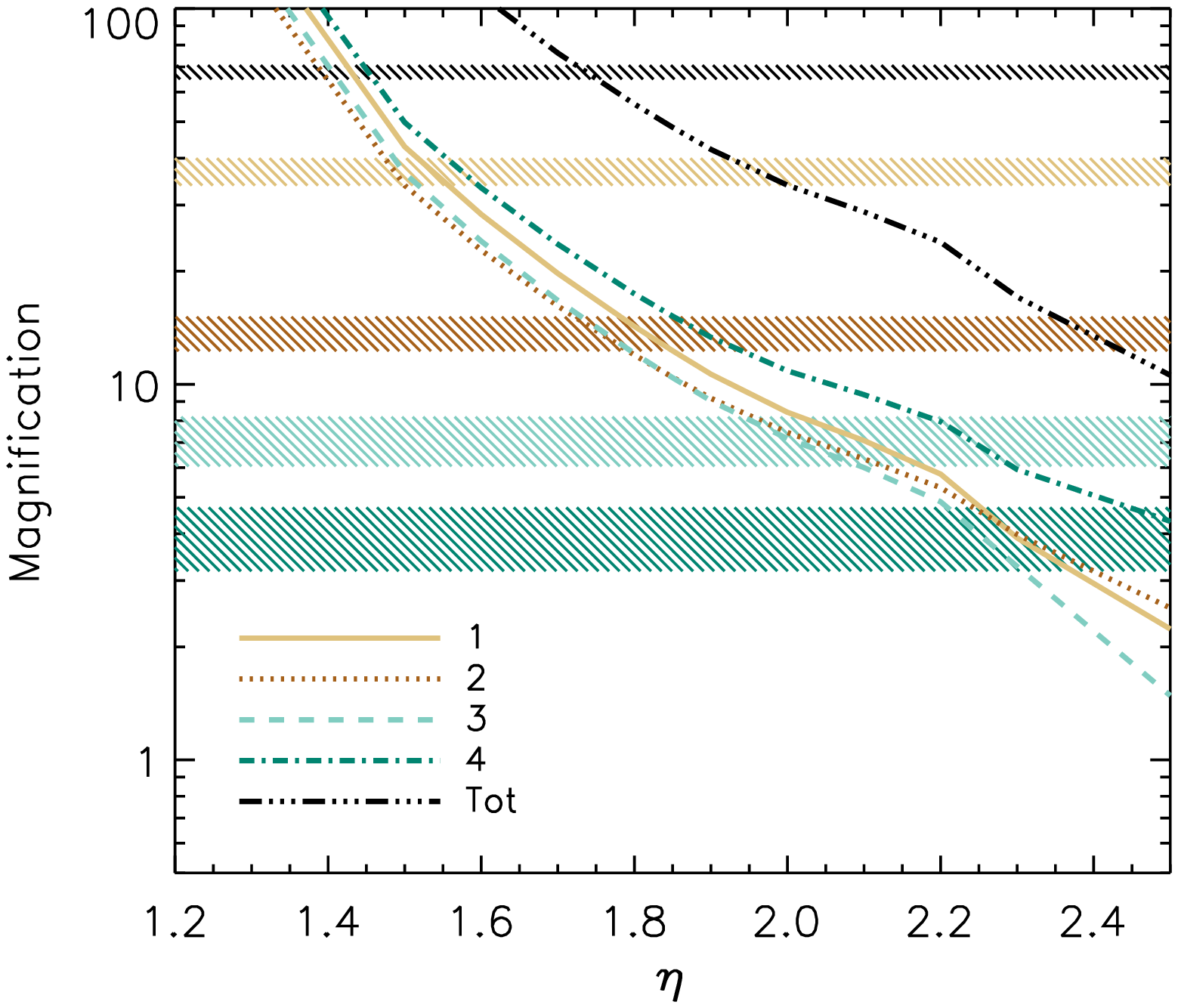}
	\caption{The predicted magnifications for the \geu images 1-4 and their total magnification, as a function of the lens mass slope $\eta$. The value $\eta=2$ represents an isothermal halo. The horizontal bands indicate the observed magnifications listed in table~\ref{tab:SNpos}. The predicted magnifications decrease monotonically with $\eta$. In contrast to the observed values, the models always predict similar magnifications for all four images, with image 4 being the brightest.
	\label{fig:magplot}}
\end{figure}

Since the inferred value of the Hubble constant depends directly on the slope of the projected lens mass $\eta$, and smaller $\eta$ implies larger image magnifications, in this section we take advantage of the standard candle nature of \sneia observationally constraining the image magnifications as listed in table~\ref{tab:SNpos}, in order to resolve the tension between the inferred slope from HST and Keck data.   
In figure~\ref{fig:magplot}, the predicted magnifications for the individual \geu images as a function of the slope $\eta$, as well as the total magnification are plotted together with the observed values indicated by the horizontal bands. Two things are immediately apparent: Ignoring substructure lensing effects, the observed total magnification suggests that $\eta\sim 1.75$ (close to the value preferred by Keck data, see figure~\ref{fig:etaplot}, and velocity dispersion observations, see figure~\ref{fig:vdispplot}): Also, regardless of the value of $\eta$, a smooth lens density fails to explain the individual \geu image fluxes, and additional sub-structure lensing is needed. This is the subject of the upcoming section.

\section{Microlens modelling}
In order to use the \geu image magnifications when modelling the macrolens mass distribution, we also have to take into the fact that substructure in the form of e.g. stars in the lens galaxy may give rise to additional (de)magnification of the \geu images.

\subsection{Stellar microlensing}\label{sec:microlensing}
Lens galaxy stars can microlens strongly lensed compact sources, such as quasars or SNe, significantly altering their observed fluxes \citep{1979Natur.282..561C}.
Although the principles of microlensing are the same as those of galaxy-scale strong gravitational lensing (see section \ref{sec:lensing}), there are some important differences between the two scenarios. 
First, the significantly lower deflector masses involved in microlensing ($m \sim M_\odot$) lead to microimage time delays and image separations that are too small to be observed with current optical instrumentation \citep[typically of order 10 $^{-6}$ seconds and 10$^{-6}$ arcsec, respectively;][]{1996IAUS..173..279M}. Second, individual deflectors are replaced  by fields of lens stars, having complex caustic patterns that vary over spatial scales of microarcseconds. These patterns can magnify or demagnify compact sources by several magnitudes over the value expected from a smooth lens model \citep[e.g.,][]{1986A&A...166...36K,1987A&A...171...49S,1992ApJ...392..424W}.
This effect has been observed in many lensed quasars  \citep[e.g.,][]{1989AJ.....98.1989I,1995ApJ...443...18W,2013A&A...553A.120T}, and simulations indicate that it should be ubiquitous in the multiple images of strongly SNe \citep{2006ApJ...653.1391D,2018ApJ...855...22G}.
 
To determine if microlensing can explain the flux anomalies in \geu, we use a set of publicly available microlensing magnification patterns produced by the GERLUMPH project \citep{2014ApJS..211...16V}.
The maps were generated by the inverse ray-shooting method \citep{1999JCoAM.109..353W}, in which a uniform surface density of rays is followed from the observer through a random field of point-mass deflectors in the lens galaxy, and collected in pixels on the source plane. The ray count in each pixel is proportional to the lensing magnification that a source at the position of the pixel would experience.  
 
The statistical properties of the microlensing are determined by the macrolensing model convergence and shear, $\kappa$ and $\gamma$, as well as the fraction of matter in stars, $f_*$, at the location of each image. For our best fit lens mass models, images 1 and 3 correspond to saddle points of the time delay surface in equation~\ref{eq:timedelay} (for which $\det\mathcal{A}<0$) and images 2 and 4 to minima (for which  $\det\mathcal{A}>0$ and  $\tr\mathcal{A}>0$).
In the simulations, fields of stars were realized at the location of each image by  assuming a uniform deflector mass $m =  M_\odot$, \citep{2014ApJS..211...16V}.
Studies of lens galaxy star fields have established that $m = 0.3 M_\odot$ is a more representative value \citep{2010ApJ...712..658P}, but other studies indicate that the deflector mass has a negligible effect on the microlensing magnification probability distributions \citep{1992ApJ...392..424W,1995MNRAS.276..103L, 2001MNRAS.320...21W,2004ApJ...613...77S}.

\subsection{Stellar mass fractions}
With the knowledge of stellar mass-to-light ratios, $m/L$, we can estimate the stellar mass surface density at the \geu image positions from the lens light intensity. Combined with the total surface mass density as derived from the lens modelling, we obtain the stellar mass fraction, $f_*$, needed to determine the stellar microlensing probabilities for the SN images. In \citet{Kauffmann:2002pn},  $m/L$ as a function of the host galaxy magnitude are presented for 4 out of the 5 Sloan Digital Sky Survey (SDSS) band, $g, r, i$ and $z$. Since the scatter in $m/L$ 
decreases at longer wavelengths, we use lens galaxy imaging in the HST $F814W$-band, which overlaps to large extent with the SDSS $i$-band. The lens light distribution is derived using the parametric fit described in \citet{Dhawan:2019vof}.

The absolute magnitude of the lens galaxy is $k$-corrected to the $i$-band at a redshift of $z=0.1$ is $M_i=-20.4$.  Using figure~14 in \citet{Kauffmann:2002pn}, gives a median $(m/L)_i=1.70$, in solar units for which the absolute magnitude is $M_{\odot,i}=4.58$ at $z=0.1$. The 5th, 25th, 75th and 95th percentile mass-to-light ratios are given by $(m/L)_i=[0.63,\, 1.12,\, 2.34,\, 3.63]$. The corresponding stellar mass fractions, $f_*$, at the SN image positions for $\eta=1.8$ (with similar results for other values of $\eta$) are presented in table~\ref{tab:fstar}. 

\begin{table}
  \centering
  \caption{Stellar mass fractions, $f_*$, at the \geu image positions as derived from HST $F814W$ photometry approximated by the SDSS $i$-band and lens modelling assuming a lens mass slope $\eta=1.8$. The columns correspond 5th, 25th, 50th (median), 75th and 95th percentile mass fractions.}
  \label{tab:fstar}
  \begin{tabular}{|c|c|c|c|c|c|}
    \hline
    Image & $f_*$ (5\,\%) & $f_*$ (25\,\%) & $f_*$ (50\,\%) & $f_*$ (75\,\%) & $f_*$ (95\,\%)\\
    \hline
1 &  $0.073$ &  $0.130$ &  $0.198$ &  $0.271$ &  $0.422$ \\
2 &  $0.069$ &  $0.122$ &  $0.186$ &  $0.256$ &  $0.397$ \\
3 &  $0.065$ &  $0.115$ &  $0.175$ &  $0.241$ &  $0.373$ \\
4 &  $0.067$ &  $0.118$ &  $0.180$ &  $0.247$ &  $0.384$ \\
\hline
  \end{tabular}
\end{table}

\subsection{Stellar microlensing probabilities}\label{sec:microlensingprob}
\begin{figure*}
    \centering
    \includegraphics[width=1\textwidth]{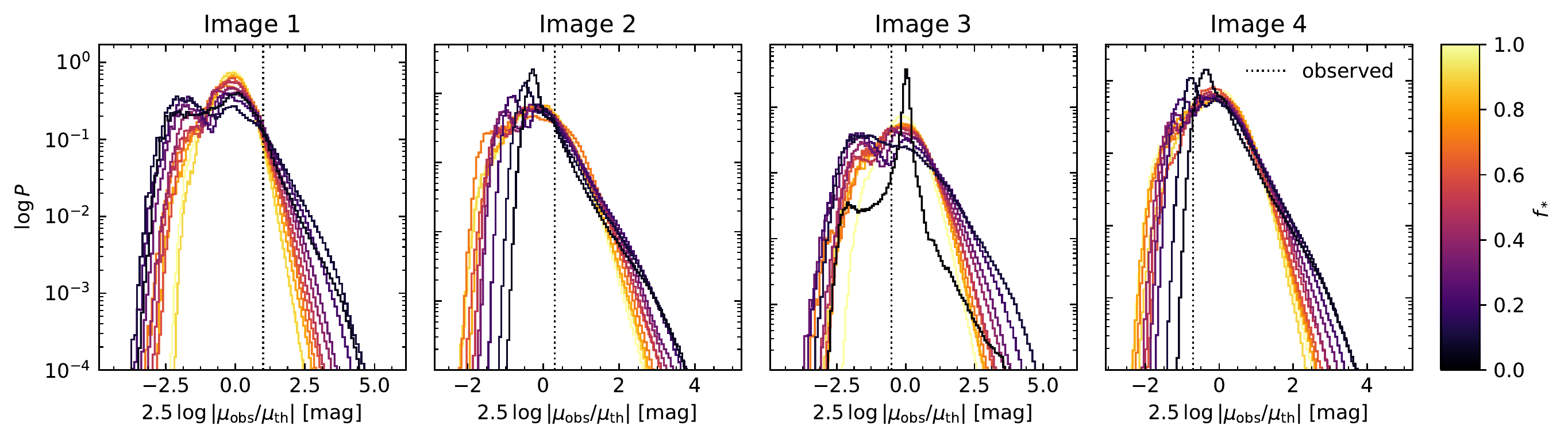}
    \caption{Microlensing magnification PDFs for each image of \geu. 
    Histograms show the distribution of differences, in magnitudes, between the total magnification of each image, $\mu_{\rm tot}$, and the magnification predicted from a smooth model, $\mu_{\rm th}$. The required microlensing magnifications as derived from the observed image fluxes are indicated by vertical dotted lines. $f_*$ denotes the fraction of matter in stars at the location of each image.}
    \label{fig:ml}
\end{figure*}

\begin{figure}
    \centering
	\includegraphics[width=\linewidth]{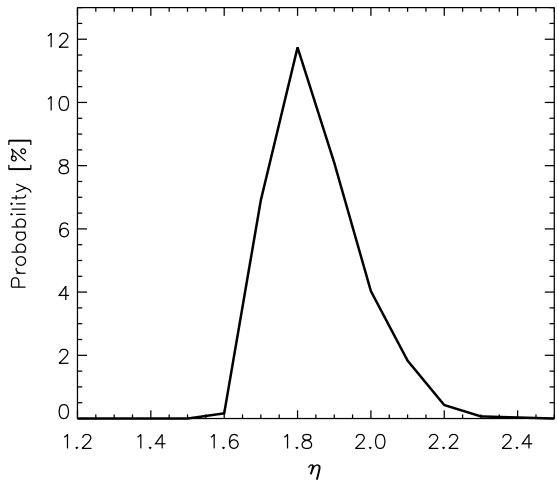}
	\caption{The probability for stellar microlensing to explain the observed \geu image fluxes as a function of the lens mass slope $\eta$, calculated in steps of $\Delta\eta = 0.1$. The probability is maximized at $12\,\%$ for $\eta=1.8$.
	\label{fig:pplot}}
\end{figure}

In figure~\ref{fig:ml}, we show the microlensing magnification probability density functions (PDFs) for each image, assuming a slope of $\eta=1.8$. Here, $\mu_{\rm tot}$ is the total magnification of each image and $\mu_{\rm th}$ the (absolute value of the) smooth lens model magnification. 
Thanks to the standard candle nature of \sneia, the required microlensing magnification can be estimated (vertical dotted lines) and compared to the microlensing PDFs, assuming different values of $\eta$.
The probability  for each image is obtained by taking a weighted average of the one-tailed $p$-value for the corresponding $f_*$ (listed in table~\ref{tab:fstar} for the case of $\eta= 1.8$). The total probability, $p_{\rm tot}$, for stellar microlensing to explain the observed \geu image fluxes is obtained by combining the individual image probabilities using the Fisher's combined probability test \citep{fisher1925statistical}. As shown in figure~\ref{fig:pplot}, this probability is maximized for $\eta=1.8$, yielding $p_{\rm tot}=0.12$, primarily driven by the large microlensing magnification required for image 1. In table~\ref{tab:ml_pvalues}, we list the convergence $\kappa$, the shear $\gamma$,  the (signed) smooth lens model  magnification $\mu_{\rm th}$, the observed magnifications $\mu_{\rm obs}$, and the probability $p$ for stellar microlensing to accommodate the difference between $\mu_{\rm th}$ and $\mu_{\rm obs}$ for the case of $\eta=1.8$. We note that the probability derived is higher than that obtained in \cite{Yahalomi:2017ihe}. This can be attributed to the difference in the assumed slope of the lens mass distributions, our updated magnifications factors taking dust extinction into account, and differences in the way the individual probabilities for each image are combined.
We conclude that stellar microlensing, with an acceptable level of probability can explain the observed flux anomalies for \geu. Note that the use of microlensing to place constraints on the macrolens model is only possible since the source is a precisely calibrated standard candle.

\begin{table}
    \centering
    \begin{tabular}{|c|c|c|c|c|}
        \hline
         Image & $\kappa$ & $\gamma$ & $\mu_{\rm th}$ & $p$ \\
        \hline
         1 & $0.644\pm 0.004$ & $0.444\pm 0.004$ & $-14.2^{+1.1}_{-1.3}$ & $0.08$\\
         2 & $0.546\pm 0.004$ & $0.349\pm 0.004$ & $11.9^{+0.9}_{-0.7}$ & $0.21$\\
         3 & $0.653\pm 0.005$ & $0.451\pm 0.005$ & $-12.0^{+1.0}_{-1.2}$& $0.43$\\
         4 & $0.566\pm 0.003$ & $0.362\pm 0.003$ & $17.4^{+1.5}_{-1.2}$ & $0.22$\\
         \hline
    \end{tabular}
    \caption{For each image, the convergence, $\kappa$, the shear, $\gamma$, and the (signed) magnification predicted for a smooth halo, $\mu_{\rm th}$, assuming a lens mass slope of $\eta=1.8$ are listed, together with 
    the one-tailed $p$-values, derived from the microlensing PDFs shown in figure \ref{fig:ml} using stellar mass fractions, $f_*$, listed in table~\ref{tab:fstar}. The total probability for the observed magnifications is $p_{\rm tot}=0.12$.}
    \label{tab:ml_pvalues}
\end{table}

\section{Lens model and Hubble constant results}
In this section, we summarize the results on the lens mass distribution, and derive limits on the Hubble constant using the observed time delay between the \geu images. 

\subsection{Lens model parameters}
Given the constraint on the lens mass slope derived from the \geu image magnifications and velocity dispersion observations, strengthened by their agreement with LGS-AO Keck data, in the following we assume a value of $\eta=1.8$.
In table~\ref{tab:geupar}, we present the resulting lens mass distribution parameters as derived from HST $F814W$ data, and Keck $J$ reference imaging. The two independent fits display a reassuring agreement, the only notable difference being that Keck data indicates a slightly higher ellipticity of the lens mass distribution.
\begin{table}
  \centering
  \caption{Derived parameters for the lens mass distribution of the \geu system assuming a lens mass slope of $\eta=1.8$ from HST $F814W$-band data when \geu is active and Keck $J$-band reference imaging. The inclination angle $\phi_{\rm mass}$ of the major axis is defined in observers coordinates, East of North (i.e. with respect to the $y$-axis in a Cartesian coordinate system).}
  \label{tab:geupar}
  \begin{tabular}{|c|c|c|}
    \hline
    Parameter & HST F814W & Keck $J$\\
    \hline
$\te$  [''] &  $0.292 \pm 0.001$  & $0.291 \pm 0.001$\\
$\phi_{\rm mass}$ [$^\circ$] & $65.5 \pm 0.3$ & $65.0\pm 0.8$ \\
$q$ & $0.88 \pm 0.01$ & $0.77\pm 0.01$ \\
\hline
  \end{tabular}
\end{table}

\begin{figure}
    \centering
	\includegraphics[width=\linewidth]{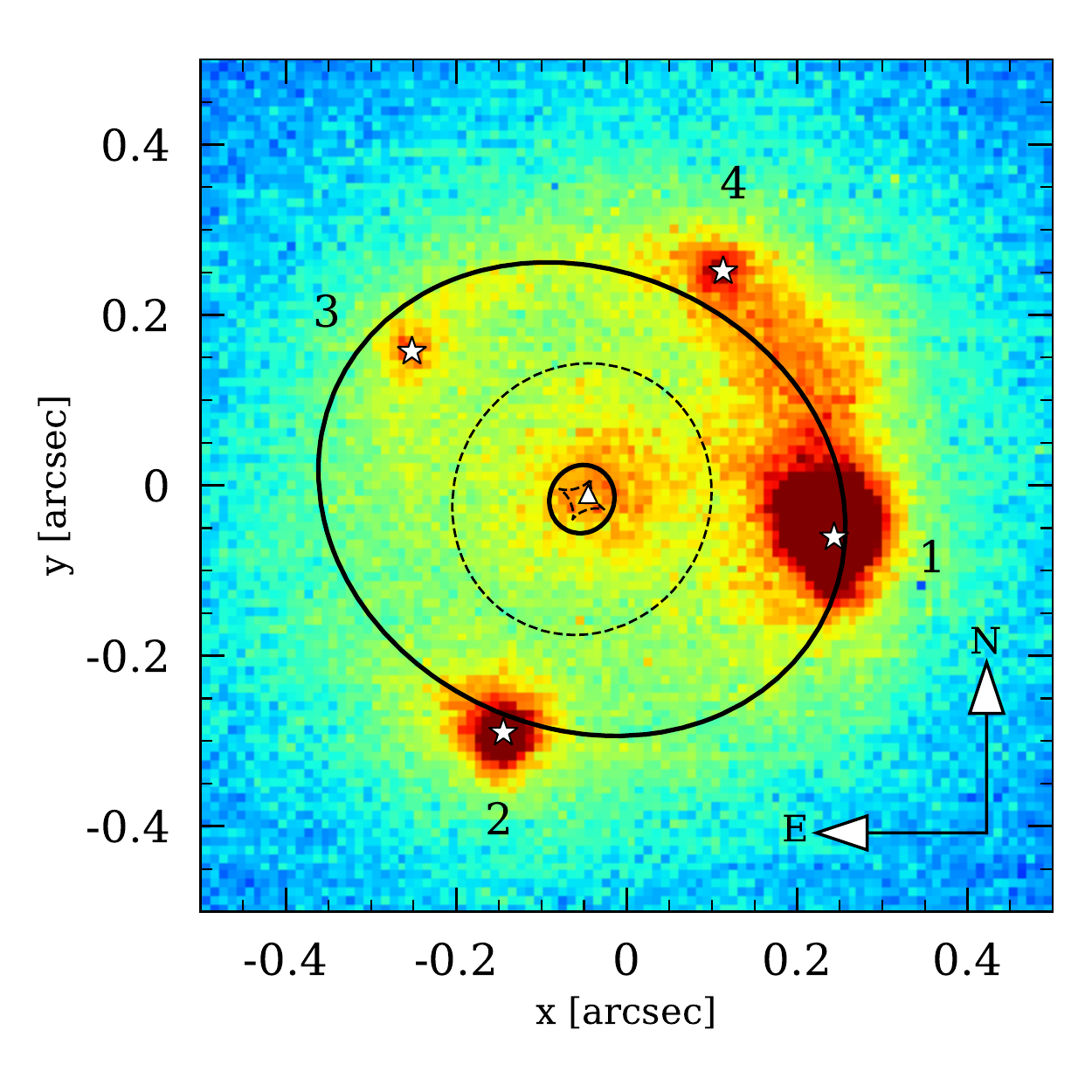}
	\caption{Critical curves (solid black) and caustics (dotted black) for the best fit lens model assuming $\eta=1.8$ overlayed on Keck LGS-AO $J$-band imaging data. The white asterisks denote \geu images positions, labelled by their number in table~\ref{tab:SNpos} \protect\citep{Goobar:2016uuf}, and the white triangle the source position. 
	\label{fig:critlines}}
\end{figure}

Figure~\ref{fig:critlines} shows the critical curves, corresponding to image positions with infinite magnification, i.e., $\det\mathcal{A} (\vec\theta)=0$, and caustics, the corresponding source positions, for the best fit lens model with slope $\eta=1.8$ overlayed on Keck LGS-AO $J$-band imaging data.  

Given the high degree of circular symmetry of the lens system, the ellipticity of the lensing galaxy is degenerate with a possible external shear component for light deflections close to the critical curve.
If the external shear is caused by a SIS galaxy, its magnitude is given by 
\be
\gamma_{\rm ext}=\frac{b_{\rm ext}}{2 r}=\frac{2\pi\sigma_v^2}{r}\left(\frac{\dls}{1157\,{\rm Mpc}}\right),
\ee 
where $b_{\rm ext}$ is the Einstein radius of the external perturber, $r$ the projected distance between the lensing galaxies and $\sigma_v$ the velocity dispersion of the perturber.
The largest expected contribution from galaxies in a square field with side length 100 arcsec centred on the \geu system is from a galaxy with $r\approx 58$ arcsec at $\phi\approx 40^\circ$ and an SDSS photometric redshift estimate of $z=0.336\pm 0.0365$ \citep{2017AJ....154...28B}. From its $g$-band magnitude of $m_g=20.77$, we can estimate its velocity dispersion to $\sigma_v\approx 187$ km/s \citep{Mitchell:2004gw,Jonsson:2008xq}. This corresponds to a shear contribution of $\gamma_{\rm ext}=1.4\cdot 10^{-4}$ for \geu. Other individual galaxies in the field contribute at most $20\,\%$ of this value and we conclude that the major part of the shear is caused by the ellipticity of the lensing galaxy itself.

\subsection{Time delays and $h$}
The measured time delays in table~\ref{tab:deltobs} can be used to constrain the value of the time delay distance $D_{\Delta t}$, using equation~\ref{eq:timedelay}, and finally the Hubble constant $h$. In practice, this is done using {\tt lenstronomy} in a manner very similar to when constraining only the lens and source properties. The only difference is that we also include the time delays as observational constraints. In principle, including time delays may also change the derived lens properties. However, since in the case of \geu the time delay uncertainties are large, in practice, they will only constrain $D_{\Delta t}$.

\begin{figure*}
   \centering
	\includegraphics[width=0.7\linewidth]{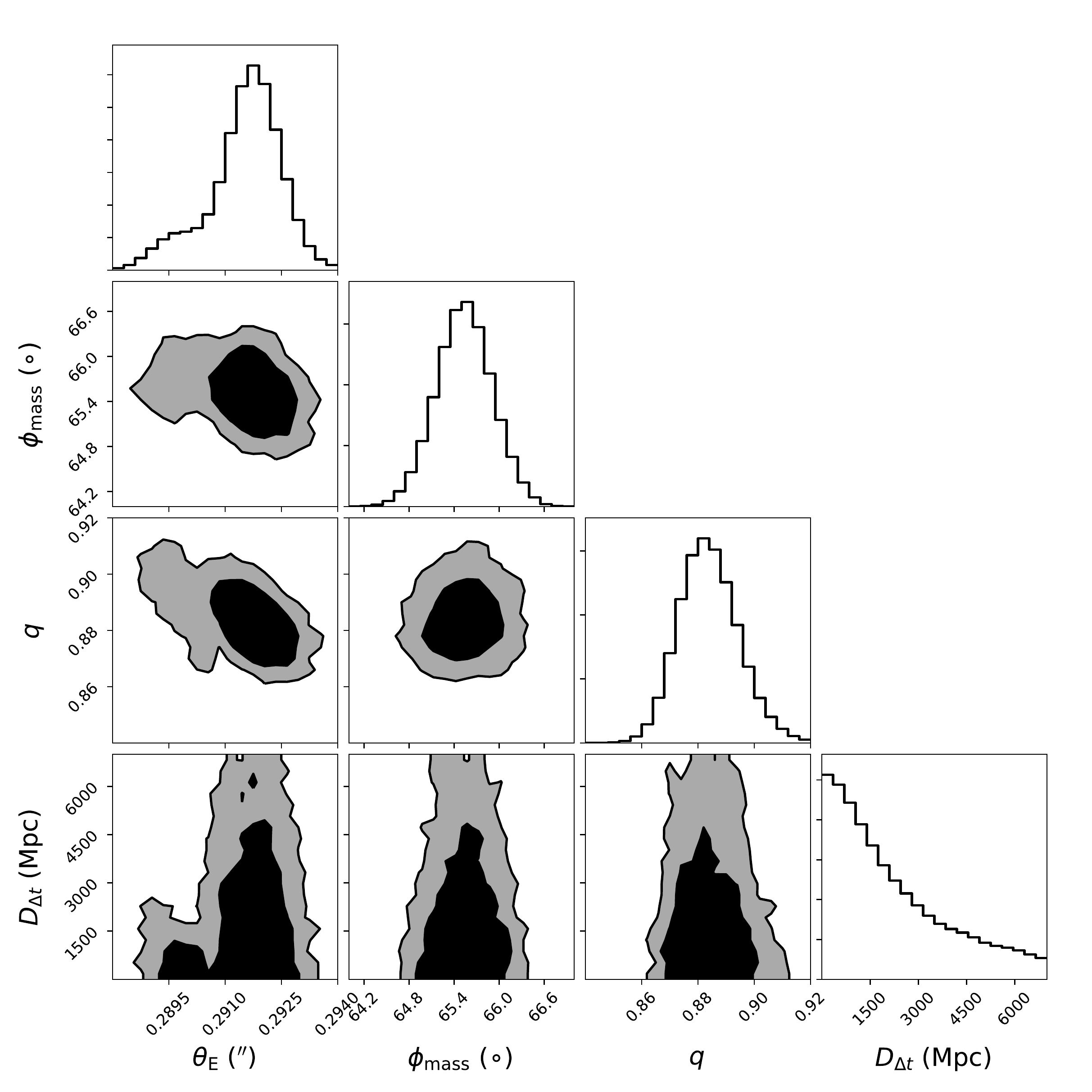}
	\caption{Confidence contours for the time delay distance $D_{\Delta t}$, together with the lens mass parameters $\te$, $\phi_{\rm mass}$ and $q$, derived from HST $F814W$-data assuming a lens mass slope of $\eta =1.8$. The corresponding limits on the Hubble constant are $h>0.5$ and $h>0.2$, at 68.3\,\% and 95\,\% CL, respectively. \label{fig:F814_SN_g18_Ddt_list}}
\end{figure*}

Given the large uncertainties in $D_{\Delta t}$, we have assumed fixed (Planck) values for cosmological parameters other than $h$ when translating from $D_{\Delta t}$ to $h$. Figure~\ref{fig:F814_SN_g18_Ddt_list}, shows 
the confidence contours for $D_{\Delta t}$, together with the lens mass parameters $\te$, $\phi_{\rm mass}$ and $q$, derived from HST $F814W$-data assuming a lens mass slope of $\eta =1.8$.
Since the observed time delays are consistent with being zero at $1\,\sigma$, we can only (weakly) constrain $D_{\Delta t}$ from above, and consequently $h$ from below to $h>0.5$ and $h>0.2$ at $68.3\,\%$ and $95\,\%$ CL, respectively. 
Since the derived Hubble constant scales with the observed time delays, $\Delta t_{ij}$, and the slope of the lens mass distribution as $h\propto (\eta-1)/\Delta t_{ij}$, conservatively assuming a very shallow lens profile with $\eta=1.6$, the constraints are slightly weakened to $h>0.4$ and $h>0.15$ at $68.3\,\%$ and $95\,\%$ CL.

As discussed in \cite{Dhawan:2019vof}, we expect to be able to decrease the uncertainty on $h$ considerably, given a strongly lensed \snia observed closer to its maximum light, and with larger time delays between the images.

\section{Conclusions}
With the aid of follow-up observations of \geu, we derive tighter constraints on the lens model, as well as the first constraint on the Hubble constant from observations of a strongly lensed \snia, following Refsdal's original proposal \citep{1964MNRAS.128..307R}. The projected mass distribution of the lens galaxy is determined, including the slope which is constrained by the \geu image magnifications and velocity dispersion measurements to $\eta\sim 1.8$,  flatter than an isothermal profile for which $\eta=2$. 

The fact that \geu exploded very close to the centre of the inner caustic of the lens, makes the predicted images for a smooth lens to be similar, in terms of radial position, magnification and arrival time. 
Since the majority of the data is post first maximum, the measurement of the time delays between images is challenging. Combined with the fact that the absolute time delays are small, the resulting fractional uncertainties are large. We expect constraints on the Hubble constant to be dominated by uncertainties in the arrival time of the images and obtain a relatively weak lower limit on the current expansion rate of $h\gtrsim 0.4$ at $68.3\,\%$ confidence level.

Observational limits on the image fluxes are better constrained, and show beyond doubt that substantial additional (de)magnification from substructure, possibly stars, in the lens has to take place. Compared to macrolens predictions, images 1 and 2 need an additional magnification of $\Delta m_1 = -1.0$ and $\Delta m_2 = -0.3$, whereas images 3 and 4 require a demagnification of $\Delta m_3 = 0.5$ and $\Delta m_4 = 0.7$, possibly from substructure lensing.
Estimating the stellar mass fractions at the \geu image positions, we derive stellar microlensing magnification probabilities for each image. The total probability for stellar microlensing to explain the observed flux anomalies
is $p_{\rm tot}\sim 12\,\%$. We conclude that the \geu flux "anomalies" are within stellar microlensing predictions.
This is the first time that microlensing has been used in combination with a precisely
calibrated standard candle to place constraints on the macrolens model in a strong lensing system.

\section*{Acknowledgements}
We thank the anonymous referee for many insightful comments helping to improve the quality of the manuscript. Also, thanks to Simon Birrer, Rahul Gupta and Mattia Bulla for help running {\tt lenstronomy}. AG acknowledges support from the Swedish National Space Agency and the Swedish Research Council.





\bibliographystyle{mnras}
\bibliography{iPTFgeu_H0} 

\begin{thebibliography}{}
\makeatletter
\relax
\def\mn@urlcharsother{\let\do\@makeother \do\$\do\&\do\#\do\^\do\_\do\%\do\~}
\def\mn@doi{\begingroup\mn@urlcharsother \@ifnextchar [ {\mn@doi@}
  {\mn@doi@[]}}
\def\mn@doi@[#1]#2{\def\@tempa{#1}\ifx\@tempa\@empty \href
  {http://dx.doi.org/#2} {doi:#2}\else \href {http://dx.doi.org/#2} {#1}\fi
  \endgroup}
\def\mn@eprint#1#2{\mn@eprint@#1:#2::\@nil}
\def\mn@eprint@arXiv#1{\href {http://arxiv.org/abs/#1} {{\tt arXiv:#1}}}
\def\mn@eprint@dblp#1{\href {http://dblp.uni-trier.de/rec/bibtex/#1.xml}
  {dblp:#1}}
\def\mn@eprint@#1:#2:#3:#4\@nil{\def\@tempa {#1}\def\@tempb {#2}\def\@tempc
  {#3}\ifx \@tempc \@empty \let \@tempc \@tempb \let \@tempb \@tempa \fi \ifx
  \@tempb \@empty \def\@tempb {arXiv}\fi \@ifundefined
  {mn@eprint@\@tempb}{\@tempb:\@tempc}{\expandafter \expandafter \csname
  mn@eprint@\@tempb\endcsname \expandafter{\@tempc}}}

\bibitem[\protect\citeauthoryear{Ade et~al.}{Ade et~al.}{2016}]{Ade:2015xua}
Ade P. A.~R.,  et~al., 2016, \mn@doi [Astron. Astrophys.]
  {10.1051/0004-6361/201525830}, 594, A13

\bibitem[\protect\citeauthoryear{Aghanim et~al.}{Aghanim
  et~al.}{2016}]{Aghanim:2016yuo}
Aghanim N.,  et~al., 2016, \mn@doi [Astron. Astrophys.]
  {10.1051/0004-6361/201628890}, 596, A107

\bibitem[\protect\citeauthoryear{Amanullah, M{\"o}rtsell  \& Goobar}{Amanullah
  et~al.}{2003}]{Amanullah:2002xh}
Amanullah R.,  M{\"o}rtsell E.,   Goobar A.,  2003, \mn@doi [Astron.
  Astrophys.] {10.1051/0004-6361:20021547}, 397, 819

\bibitem[\protect\citeauthoryear{{Amanullah} et~al.,}{{Amanullah}
  et~al.}{2011}]{2011ApJ...742L...7A}
{Amanullah} R.,  et~al., 2011, \mn@doi [\apj] {10.1088/2041-8205/742/1/L7},
  \href {https://ui.adsabs.harvard.edu/#abs/2011ApJ...742L...7A} {742, L7}

\bibitem[\protect\citeauthoryear{Birrer \& Amara}{Birrer \&
  Amara}{2018}]{Birrer:2018xgm}
Birrer S.,  Amara A.,  2018, ] {10.1016/j.dark.2018.11.002}

\bibitem[\protect\citeauthoryear{{Blanton} et~al.,}{{Blanton}
  et~al.}{2017}]{2017AJ....154...28B}
{Blanton} M.~R.,  et~al., 2017, \mn@doi [\aj] {10.3847/1538-3881/aa7567}, \href
  {https://ui.adsabs.harvard.edu/abs/2017AJ....154...28B} {154, 28}

\bibitem[\protect\citeauthoryear{{Chae}, {Bernardi}  \& {Sheth}}{{Chae}
  et~al.}{2019}]{2019ApJ...874...41C}
{Chae} K.-H.,  {Bernardi} M.,   {Sheth} R.~K.,  2019, \mn@doi [\apj]
  {10.3847/1538-4357/ab09fd}, \href
  {https://ui.adsabs.harvard.edu/abs/2019ApJ...874...41C} {874, 41}

\bibitem[\protect\citeauthoryear{{Chang} \& {Refsdal}}{{Chang} \&
  {Refsdal}}{1979}]{1979Natur.282..561C}
{Chang} K.,  {Refsdal} S.,  1979, \mn@doi [\nat] {10.1038/282561a0}, \href
  {http://adsabs.harvard.edu/abs/1979Natur.282..561C} {282, 561}

\bibitem[\protect\citeauthoryear{Chornock et~al.}{Chornock
  et~al.}{2013}]{Chornock:2013wj}
Chornock R.,  et~al., 2013, \mn@doi [Astrophys. J.]
  {10.1088/0004-637X/767/2/162}, 767, 162

\bibitem[\protect\citeauthoryear{Dhawan, Goobar  \& M{\"o}rtsell}{Dhawan
  et~al.}{2018}]{Dhawan:2017kft}
Dhawan S.,  Goobar A.,   M{\"o}rtsell E.,  2018, \mn@doi [JCAP]
  {10.1088/1475-7516/2018/07/024}, 1807, 024

\bibitem[\protect\citeauthoryear{Dhawan et~al.}{Dhawan
  et~al.}{2020}]{Dhawan:2019vof}
Dhawan S.,  et~al., 2020, \mn@doi [Mon. Not. Roy. Astron. Soc.]
  {10.1093/mnras/stz2965}, 491, 2639

\bibitem[\protect\citeauthoryear{{Dobler} \& {Keeton}}{{Dobler} \&
  {Keeton}}{2006}]{2006ApJ...653.1391D}
{Dobler} G.,  {Keeton} C.~R.,  2006, \mn@doi [\apj] {10.1086/508769}, \href
  {http://adsabs.harvard.edu/abs/2006ApJ...653.1391D} {653, 1391}

\bibitem[\protect\citeauthoryear{{Falco}, {Gorenstein}  \& {Shapiro}}{{Falco}
  et~al.}{1985}]{1985ApJ...289L...1F}
{Falco} E.~E.,  {Gorenstein} M.~V.,   {Shapiro} I.~I.,  1985, \mn@doi [\apjl]
  {10.1086/184422}, \href
  {https://ui.adsabs.harvard.edu/abs/1985ApJ...289L...1F} {289, L1}

\bibitem[\protect\citeauthoryear{Fisher}{Fisher}{1925}]{fisher1925statistical}
Fisher R.,  1925, Statistical methods for research workers.
Edinburgh Oliver \& Boyd

\bibitem[\protect\citeauthoryear{Gerhard, Kronawitter, Saglia  \&
  Bender}{Gerhard et~al.}{2001}]{Gerhard:2000ck}
Gerhard O.,  Kronawitter A.,  Saglia R.~P.,   Bender R.,  2001, \mn@doi
  [Astron. J.] {10.1086/319940}, 121, 1936

\bibitem[\protect\citeauthoryear{{Goldstein}, {Nugent}, {Kasen}  \&
  {Collett}}{{Goldstein} et~al.}{2018}]{2018ApJ...855...22G}
{Goldstein} D.~A.,  {Nugent} P.~E.,  {Kasen} D.~N.,   {Collett} T.~E.,  2018,
  \mn@doi [\apj] {10.3847/1538-4357/aaa975}, \href
  {http://adsabs.harvard.edu/abs/2018ApJ...855...22G} {855, 22}

\bibitem[\protect\citeauthoryear{Goliath \& M{\"o}rtsell}{Goliath \&
  M{\"o}rtsell}{2000}]{Goliath:1999kpa}
Goliath M.,  M{\"o}rtsell E.,  2000, \mn@doi [Phys. Lett.]
  {10.1016/S0370-2693(00)00763-2}, B486, 249

\bibitem[\protect\citeauthoryear{{Goobar} \& {Leibundgut}}{{Goobar} \&
  {Leibundgut}}{2011}]{2011ARNPS..61..251G}
{Goobar} A.,  {Leibundgut} B.,  2011, \mn@doi [Annual Review of Nuclear and
  Particle Science] {10.1146/annurev-nucl-102010-130434}, \href
  {https://ui.adsabs.harvard.edu/\#abs/2011ARNPS..61..251G} {61, 251}

\bibitem[\protect\citeauthoryear{Goobar, M{\"o}rtsell, Amanullah  \&
  Nugent}{Goobar et~al.}{2002}]{Goobar:2002pz}
Goobar A.,  M{\"o}rtsell E.,  Amanullah R.,   Nugent P.,  2002, \mn@doi
  [Astron. Astrophys.] {10.1051/0004-6361:20020987}, 393, 25

\bibitem[\protect\citeauthoryear{{Goobar} et~al.,}{{Goobar}
  et~al.}{2009}]{2009A&A...507...71G}
{Goobar} A.,  et~al., 2009, \mn@doi [\aap] {10.1051/0004-6361/200811254}, \href
  {https://ui.adsabs.harvard.edu/abs/2009A&A...507...71G} {507, 71}

\bibitem[\protect\citeauthoryear{Goobar et~al.}{Goobar
  et~al.}{2017}]{Goobar:2016uuf}
Goobar A.,  et~al., 2017, \mn@doi [Science] {10.1126/science.aal2729}, 356, 291

\bibitem[\protect\citeauthoryear{{Hsiao}, {Conley}, {Howell}, {Sullivan},
  {Pritchet}, {Carlberg}, {Nugent}  \& {Phillips}}{{Hsiao}
  et~al.}{2007}]{2007ApJ...663.1187H}
{Hsiao} E.~Y.,  {Conley} A.,  {Howell} D.~A.,  {Sullivan} M.,  {Pritchet}
  C.~J.,  {Carlberg} R.~G.,  {Nugent} P.~E.,   {Phillips} M.~M.,  2007, \mn@doi
  [\apj] {10.1086/518232}, \href
  {https://ui.adsabs.harvard.edu/#abs/2007ApJ...663.1187H} {663, 1187}

\bibitem[\protect\citeauthoryear{{Irwin}, {Webster}, {Hewett}, {Corrigan}  \&
  {Jedrzejewski}}{{Irwin} et~al.}{1989}]{1989AJ.....98.1989I}
{Irwin} M.~J.,  {Webster} R.~L.,  {Hewett} P.~C.,  {Corrigan} R.~T.,
  {Jedrzejewski} R.~I.,  1989, \mn@doi [\aj] {10.1086/115272}, \href
  {https://ui.adsabs.harvard.edu/abs/1989AJ.....98.1989I} {98, 1989}

\bibitem[\protect\citeauthoryear{{Johansson} et~al.,}{{Johansson}
  et~al.}{2020}]{2020arXiv200410164J}
{Johansson} J.,  et~al., 2020, arXiv e-prints, \href
  {https://ui.adsabs.harvard.edu/abs/2020arXiv200410164J} {p. arXiv:2004.10164}

\bibitem[\protect\citeauthoryear{J{\"o}nsson, Dahl{\'e}n, Goobar, Gunnarsson,
  M{\"o}rtsell  \& Lee}{J{\"o}nsson et~al.}{2006}]{Jonsson:2005qv}
J{\"o}nsson J.,  Dahl{\'e}n T.,  Goobar A.,  Gunnarsson C.,  M{\"o}rtsell E.,
  Lee K.,  2006, \mn@doi [Astrophys. J.] {10.1086/499396}, 639, 991

\bibitem[\protect\citeauthoryear{J{\"o}nsson, Dahl{\'e}n, Goobar, M{\"o}rtsell
  \& Riess}{J{\"o}nsson et~al.}{2007}]{Jonsson:2006eu}
J{\"o}nsson J.,  Dahl{\'e}n T.,  Goobar A.,  M{\"o}rtsell E.,   Riess A.,
  2007, \mn@doi [JCAP] {10.1088/1475-7516/2007/06/002}, 0706, 002

\bibitem[\protect\citeauthoryear{J{\"o}nsson, Kronborg, M{\"o}rtsell  \&
  Sollerman}{J{\"o}nsson et~al.}{2008}]{Jonsson:2008xq}
J{\"o}nsson J.,  Kronborg T.,  M{\"o}rtsell E.,   Sollerman J.,  2008, \mn@doi
  [Astron. Astrophys.] {10.1051/0004-6361:200809729}, 487, 467

\bibitem[\protect\citeauthoryear{J{\"o}nsson, M{\"o}rtsell  \&
  Sollerman}{J{\"o}nsson et~al.}{2009}]{Jonsson:2008if}
J{\"o}nsson J.,  M{\"o}rtsell E.,   Sollerman J.,  2009, \mn@doi [Astron.
  Astrophys.] {10.1051/0004-6361:200811040}, 493, 331

\bibitem[\protect\citeauthoryear{{J{\"o}nsson}, {Dahl{\'e}n}, {Hook}, {Goobar}
  \& {M{\"o}rtsell}}{{J{\"o}nsson} et~al.}{2010}]{2010MNRAS.402..526J}
{J{\"o}nsson} J.,  {Dahl{\'e}n} T.,  {Hook} I.,  {Goobar} A.,   {M{\"o}rtsell}
  E.,  2010, \mn@doi [\mnras] {10.1111/j.1365-2966.2009.15907.x}, \href
  {https://ui.adsabs.harvard.edu/#abs/2010MNRAS.402..526J} {402, 526}

\bibitem[\protect\citeauthoryear{{Kassiola} \& {Kovner}}{{Kassiola} \&
  {Kovner}}{1993}]{1993LIACo..31..571K}
{Kassiola} A.,  {Kovner} I.,  1993, in {Surdej} J.,  {Fraipont-Caro} D.,
  {Gosset} E.,  {Refsdal} S.,   {Remy} M.,  eds,  Liege International
  Astrophysical Colloquia Vol. 31, Liege International Astrophysical Colloquia.
  p.~571

\bibitem[\protect\citeauthoryear{Kauffmann et~al.}{Kauffmann
  et~al.}{2003}]{Kauffmann:2002pn}
Kauffmann G.,  et~al., 2003, \mn@doi [Mon. Not. Roy. Astron. Soc.]
  {10.1046/j.1365-8711.2003.06291.x}, 341, 33

\bibitem[\protect\citeauthoryear{{Kayser}, {Refsdal}  \& {Stabell}}{{Kayser}
  et~al.}{1986}]{1986A&A...166...36K}
{Kayser} R.,  {Refsdal} S.,   {Stabell} R.,  1986, \aap, \href
  {https://ui.adsabs.harvard.edu/abs/1986A%26A...166...36K} {166, 36}

\bibitem[\protect\citeauthoryear{Keeton}{Keeton}{2001b}]{Keeton:2001ss}
Keeton C.~R.,  2001b

\bibitem[\protect\citeauthoryear{Keeton}{Keeton}{2001a}]{Keeton:2001sr}
Keeton C.~R.,  2001a

\bibitem[\protect\citeauthoryear{Kelly et~al.}{Kelly
  et~al.}{2016a}]{Kelly:2015xvu}
Kelly P.~L.,  et~al., 2016a, \mn@doi [Astrophys. J.]
  {10.3847/2041-8205/819/1/L8}, 819, L8

\bibitem[\protect\citeauthoryear{Kelly et~al.}{Kelly
  et~al.}{2016b}]{Kelly:2015vjq}
Kelly P.~L.,  et~al., 2016b, \mn@doi [Astrophys. J.]
  {10.3847/0004-637X/831/2/205}, 831, 205

\bibitem[\protect\citeauthoryear{Kolatt \& Bartelmann}{Kolatt \&
  Bartelmann}{1998}]{Kolatt:1997zh}
Kolatt T.~S.,  Bartelmann M.,  1998, \mn@doi [Mon. Not. Roy. Astron. Soc.]
  {10.1046/j.1365-8711.1998.01466.x}, 296, 763

\bibitem[\protect\citeauthoryear{{Kormann}, {Schneider}  \&
  {Bartelmann}}{{Kormann} et~al.}{1994}]{1994A&A...284..285K}
{Kormann} R.,  {Schneider} P.,   {Bartelmann} M.,  1994, \aap, \href
  {http://adsabs.harvard.edu/abs/1994A%26A...284..285K} {284, 285}

\bibitem[\protect\citeauthoryear{{Lewis} \& {Irwin}}{{Lewis} \&
  {Irwin}}{1995}]{1995MNRAS.276..103L}
{Lewis} G.~F.,  {Irwin} M.~J.,  1995, \mn@doi [\mnras]
  {10.1093/mnras/276.1.103}, \href
  {https://ui.adsabs.harvard.edu/abs/1995MNRAS.276..103L} {276, 103}

\bibitem[\protect\citeauthoryear{Metcalf \& Silk}{Metcalf \&
  Silk}{1999}]{Metcalf:1999qb}
Metcalf R.~B.,  Silk J.,  1999, \mn@doi [Astrophys. J.] {10.1086/312086}, 519,
  L1

\bibitem[\protect\citeauthoryear{Mitchell, Keeton, Frieman  \& Sheth}{Mitchell
  et~al.}{2005}]{Mitchell:2004gw}
Mitchell J.~L.,  Keeton C.~R.,  Frieman J.~A.,   Sheth R.~K.,  2005, \mn@doi
  [Astrophys. J.] {10.1086/427910}, 622, 81

\bibitem[\protect\citeauthoryear{{Moore} \& {Hewitt}}{{Moore} \&
  {Hewitt}}{1996}]{1996IAUS..173..279M}
{Moore} C.~B.,  {Hewitt} J.~N.,  1996, in {Kochanek} C.~S.,  {Hewitt} J.~N.,
  eds,  IAU Symposium Vol. 173, Astrophysical Applications of Gravitational
  Lensing. p.~279

\bibitem[\protect\citeauthoryear{More, Suyu, Oguri, More  \& Lee}{More
  et~al.}{2017}]{More:2016sys}
More A.,  Suyu S.~H.,  Oguri M.,  More S.,   Lee C.-H.,  2017, \mn@doi
  [Astrophys. J.] {10.3847/2041-8213/835/2/L25}, 835, L25

\bibitem[\protect\citeauthoryear{M{\"o}rtsell}{M{\"o}rtsell}{2002}]{Mortsell:2001es}
M{\"o}rtsell E.,  2002, \mn@doi [Astron. Astrophys.]
  {10.1051/0004-6361:20011653}, 382, 787

\bibitem[\protect\citeauthoryear{M{\"o}rtsell \& Dhawan}{M{\"o}rtsell \&
  Dhawan}{2018}]{Mortsell:2018mfj}
M{\"o}rtsell E.,  Dhawan S.,  2018, \mn@doi [JCAP]
  {10.1088/1475-7516/2018/09/025}, 1809, 025

\bibitem[\protect\citeauthoryear{M{\"o}rtsell \& Sunesson}{M{\"o}rtsell \&
  Sunesson}{2006}]{Mortsell:2005mf}
M{\"o}rtsell E.,  Sunesson C.,  2006, \mn@doi [JCAP]
  {10.1088/1475-7516/2006/01/012}, 0601, 012

\bibitem[\protect\citeauthoryear{M{\"o}rtsell, Goobar  \&
  Bergstrom}{M{\"o}rtsell et~al.}{2001a}]{Moertsell:2001ah}
M{\"o}rtsell E.,  Goobar A.,   Bergstrom L.,  2001a, \mn@doi [Astrophys. J.]
  {10.1086/322396}, 559, 53

\bibitem[\protect\citeauthoryear{M{\"o}rtsell, Gunnarsson  \&
  Goobar}{M{\"o}rtsell et~al.}{2001b}]{Mortsell:2001rb}
M{\"o}rtsell E.,  Gunnarsson C.,   Goobar A.,  2001b, \mn@doi [Astrophys. J.]
  {10.1086/323242}, 561, 106

\bibitem[\protect\citeauthoryear{M{\"o}rtsell, Dahle  \&
  Hannestad}{M{\"o}rtsell et~al.}{2005}]{Mortsell:2004py}
M{\"o}rtsell E.,  Dahle H.,   Hannestad S.,  2005, \mn@doi [Astrophys. J.]
  {10.1086/426706}, 619, 733

\bibitem[\protect\citeauthoryear{Nordin et~al.}{Nordin
  et~al.}{2014}]{Nordin:2013cfa}
Nordin J.,  et~al., 2014, \mn@doi [Mon. Not. Roy. Astron. Soc.]
  {10.1093/mnras/stu376}, 440, 2742

\bibitem[\protect\citeauthoryear{Oguri \& Kawano}{Oguri \&
  Kawano}{2003}]{Oguri:2002ku}
Oguri M.,  Kawano Y.,  2003, \mn@doi [Mon. Not. Roy. Astron. Soc.]
  {10.1046/j.1365-8711.2003.06290.x}, 338, L25

\bibitem[\protect\citeauthoryear{{Petrushevska} et~al.,}{{Petrushevska}
  et~al.}{2016}]{2016A&A...594A..54P}
{Petrushevska} T.,  et~al., 2016, \mn@doi [\aap] {10.1051/0004-6361/201628925},
  \href {https://ui.adsabs.harvard.edu/abs/2016A&A...594A..54P} {594, A54}

\bibitem[\protect\citeauthoryear{{Poindexter} \& {Kochanek}}{{Poindexter} \&
  {Kochanek}}{2010}]{2010ApJ...712..658P}
{Poindexter} S.,  {Kochanek} C.~S.,  2010, \mn@doi [\apj]
  {10.1088/0004-637X/712/1/658}, \href
  {https://ui.adsabs.harvard.edu/abs/2010ApJ...712..658P} {712, 658}

\bibitem[\protect\citeauthoryear{Quimby et~al.}{Quimby
  et~al.}{2013}]{Quimby:2013lfa}
Quimby R.~M.,  et~al., 2013, \mn@doi [Astrophys. J.]
  {10.1088/2041-8205/768/1/L20}, 768, L20

\bibitem[\protect\citeauthoryear{{Rauch}}{{Rauch}}{1991}]{1991ApJ...374...83R}
{Rauch} K.~P.,  1991, \mn@doi [\apj] {10.1086/170098}, \href
  {https://ui.adsabs.harvard.edu/abs/1991ApJ...374...83R} {374, 83}

\bibitem[\protect\citeauthoryear{{Refsdal}}{{Refsdal}}{1964}]{1964MNRAS.128..307R}
{Refsdal} S.,  1964, \mn@doi [\mnras] {10.1093/mnras/128.4.307}, \href
  {http://adsabs.harvard.edu/abs/1964MNRAS.128..307R} {128, 307}

\bibitem[\protect\citeauthoryear{Riess, Casertano, Yuan, Macri  \&
  Scolnic}{Riess et~al.}{2019}]{Riess:2019cxk}
Riess A.~G.,  Casertano S.,  Yuan W.,  Macri L.~M.,   Scolnic D.,  2019

\bibitem[\protect\citeauthoryear{Rodney et~al.}{Rodney
  et~al.}{2015}]{Rodney:2015lpa}
Rodney S.~A.,  et~al., 2015, \mn@doi [Astrophys. J.]
  {10.1088/0004-637X/811/1/70}, 811, 70

\bibitem[\protect\citeauthoryear{Rubin et~al.}{Rubin
  et~al.}{2018}]{Rubin:2017ipu}
Rubin D.,  et~al., 2018, \mn@doi [Astrophys. J.] {10.3847/1538-4357/aad565},
  866, 65

\bibitem[\protect\citeauthoryear{{Schechter}, {Wambsganss}  \&
  {Lewis}}{{Schechter} et~al.}{2004}]{2004ApJ...613...77S}
{Schechter} P.~L.,  {Wambsganss} J.,   {Lewis} G.~F.,  2004, \mn@doi [\apj]
  {10.1086/422907}, \href
  {https://ui.adsabs.harvard.edu/abs/2004ApJ...613...77S} {613, 77}

\bibitem[\protect\citeauthoryear{{Schneider} \& {Weiss}}{{Schneider} \&
  {Weiss}}{1987}]{1987A&A...171...49S}
{Schneider} P.,  {Weiss} A.,  1987, \aap, \href
  {https://ui.adsabs.harvard.edu/abs/1987A%26A...171...49S} {171, 49}

\bibitem[\protect\citeauthoryear{Seljak \& Holz}{Seljak \&
  Holz}{1999}]{Seljak:1999tm}
Seljak U.,  Holz D.~E.,  1999, Astron. Astrophys., 351, L10

\bibitem[\protect\citeauthoryear{{Tewes}, {Courbin}  \& {Meylan}}{{Tewes}
  et~al.}{2013}]{2013A&A...553A.120T}
{Tewes} M.,  {Courbin} F.,   {Meylan} G.,  2013, \mn@doi [\aap]
  {10.1051/0004-6361/201220123}, \href
  {https://ui.adsabs.harvard.edu/abs/2013A%26A...553A.120T} {553, A120}

\bibitem[\protect\citeauthoryear{{Vernardos}, {Fluke}, {Bate}  \&
  {Croton}}{{Vernardos} et~al.}{2014}]{2014ApJS..211...16V}
{Vernardos} G.,  {Fluke} C.~J.,  {Bate} N.~F.,   {Croton} D.,  2014, \mn@doi
  [\apjs] {10.1088/0067-0049/211/1/16}, \href
  {https://ui.adsabs.harvard.edu/abs/2014ApJS..211...16V} {211, 16}

\bibitem[\protect\citeauthoryear{{Wambsganss}}{{Wambsganss}}{1992}]{1992ApJ...392..424W}
{Wambsganss} J.,  1992, \mn@doi [\apj] {10.1086/171441}, \href
  {https://ui.adsabs.harvard.edu/abs/1992ApJ...392..424W} {392, 424}

\bibitem[\protect\citeauthoryear{{Wambsganss}}{{Wambsganss}}{1999}]{1999JCoAM.109..353W}
{Wambsganss} J.,  1999, Journal of Computational and Applied Mathematics, \href
  {https://ui.adsabs.harvard.edu/abs/1999JCoAM.109..353W} {109, 353}

\bibitem[\protect\citeauthoryear{{Witt}, {Mao}  \& {Schechter}}{{Witt}
  et~al.}{1995}]{1995ApJ...443...18W}
{Witt} H.~J.,  {Mao} S.,   {Schechter} P.~L.,  1995, \mn@doi [\apj]
  {10.1086/175499}, \href
  {https://ui.adsabs.harvard.edu/abs/1995ApJ...443...18W} {443, 18}

\bibitem[\protect\citeauthoryear{{Wyithe} \& {Turner}}{{Wyithe} \&
  {Turner}}{2001}]{2001MNRAS.320...21W}
{Wyithe} J.~S.~B.,  {Turner} E.~L.,  2001, \mn@doi [\mnras]
  {10.1046/j.1365-8711.2001.03917.x}, \href
  {https://ui.adsabs.harvard.edu/abs/2001MNRAS.320...21W} {320, 21}

\bibitem[\protect\citeauthoryear{Yahalomi, Schechter  \& Wambsganss}{Yahalomi
  et~al.}{2017}]{Yahalomi:2017ihe}
Yahalomi D.~A.,  Schechter P.~L.,   Wambsganss J.,  2017

\bibitem[\protect\citeauthoryear{Zumalacarregui \& Seljak}{Zumalacarregui \&
  Seljak}{2018}]{Zumalacarregui:2017qqd}
Zumalacarregui M.,  Seljak U.,  2018, \mn@doi [Phys. Rev. Lett.]
  {10.1103/PhysRevLett.121.141101}, 121, 141101

\makeatother
\end{thebibliography}




\bsp	
\label{lastpage}
\end{document}